\documentclass[aos,MSNbibl,seceqn,nameyear,dvips]{arximspdf}
\usepackage{dcolumn}
\usepackage{graphicx}

\doi{10.1214/11-AOS944}
\volume{39}
\issue{6}
\pubyear{2011}
\firstpage{3320}
\lastpage{3356}

\makeatletter

\newcolumntype{d}[1]{D{.}{.}{#1}}

\renewcommand{\hat}{\widehat}

\newcommand{\Max}{\mathrm{MAX}}

\newcommand{\bff}{\mathbf{f}}
\newcommand{\by}{\mathbf{y}}
\newcommand{\bA}{\mathbf{A}}
\newcommand{\bu}{\mathbf{u}}
\newcommand{\bv}{\mathbf{v}}
\newcommand{\bB}{\mathbf{B}}
\newcommand{\bC}{\mathbf{C}}
\newcommand{\bD}{\mathbf{D}}
\newcommand{\bE}{\mathbf{E}}
\newcommand{\bM}{\mathbf{M}}
\newcommand{\bS}{\mathbf{S}}
\newcommand{\bX}{\mathbf{X}}
\newcommand{\bone}{\mathbf{1}}
\newcommand{\beps}{\bolds\varepsilon}
\newcommand{\bmu}{\bolds\mu}
\newcommand{\bOmega}{\bolds\Omega}
\newcommand{\bPhi}{\bolds\Phi}

\newcommand{\hB}{\widehat{\bB}}
\newcommand{\ty}{\tilde{\by}}
\newcommand{\tf}{\tilde{\bff}}
\newcommand{\tB}{\tilde{\bB}}
\newcommand{\hA}{\widehat{\bA}}
\newcommand{\hu}{\widehat{\bu}}
\newcommand{\hcov}{\operatorname{\widehat{cov}}}
\newcommand{\hSig}{\widehat{\Sig}}
\newcommand{\cov}{\operatorname{cov}}
\newcommand{\Sig}{\bolds{\Sigma}}
\newcommand{\tr}{\operatorname{tr}}
\newcommand{\bw}{\mathbf{w}}
\newcommand{\var}{\operatorname{var}}

\newtheorem{theorem}{Theorem}[section]
\newtheorem{lem}{Lemma}[section]

\newproclaim{assum}{Assumption}[section]
\newproclaim{remark}{Remark}[section]

\makeatother

\begin{document}
\begin{frontmatter}

\title{High-dimensional covariance matrix estimation in approximate
factor models\thanksref{T1}}
\runtitle{Covariance matrix in factor model}

\thankstext{T1}{Supported in part by NIH Grants R01-GM100474,
R01-GM072611 and NSF Grant DMS-07-04337.}

\begin{aug}
\author[A]{\fnms{Jianqing} \snm{Fan}\ead[label=e1]{jqfan@princeton.edu}},
\author[A]{\fnms{Yuan} \snm{Liao}\corref{}\ead[label=e2]{yuanliao@princeton.edu}}
\and
\author[A]{\fnms{Martina} \snm{Mincheva}\ead[label=e3]{mincheva@princeton.edu}}
\runauthor{J. Fan, Y. Liao and M. Mincheva}
\affiliation{Princeton University}
\address[A]{Department of Operations Research\\
\quad and Financial Engineering\\
Princeton University\\
Princeton, New Jersey 08544\\
USA\\
\printead{e1}\\
\hphantom{E-mail: }\printead*{e2}\\
\hphantom{E-mail: }\printead*{e3}} 
\end{aug}

\received{\smonth{5} \syear{2011}}
\revised{\smonth{10} \syear{2011}}

%
\begin{abstract}
The variance--covariance matrix plays a central role in the inferential
theories of high-dimensional factor models in finance and economics.
Popular regularization methods of directly exploiting sparsity are not
directly applicable to many financial problems. Classical methods of
estimating the covariance matrices are based on the strict factor
models, assuming independent idiosyncratic components. This assumption,
however, is restrictive in practical applications. By assuming sparse
error covariance matrix, we allow the presence of the cross-sectional
correlation even after taking out common factors, and it enables us to
combine the merits of both methods. We estimate the sparse covariance
using the adaptive thresholding technique as in Cai and Liu
[\textit{J. Amer. Statist. Assoc.} \textbf{106} (2011) 672--684],
taking into account the fact that direct observations of the
idiosyncratic components are unavailable. The impact of high
dimensionality on the covariance matrix estimation based on the factor
structure is then studied.
\end{abstract}

%
\begin{keyword}[class=AMS]
\kwd[Primary ]{62H25}
\kwd[; secondary ]{62F12}
\kwd{62H12}.
\end{keyword}
\begin{keyword}
\kwd{Sparse estimation}
\kwd{thresholding}
\kwd{cross-sectional correlation}
\kwd{common factors}
\kwd{idiosyncratic}
\kwd{seemingly unrelated regression}.
\end{keyword}

\end{frontmatter}

\section{Introduction}\label{sec1}

We consider a factor model defined as follows:
%
\begin{equation}
y_{it}=\mathbf{b}_i'\bff_t+u_{it},
\end{equation}
where $y_{it}$ is the observed datum for the $i$th ($i=1,\ldots,p$) asset
at time $t=1,\ldots,T$; ${\mathbf b}_i$ is a $K\times1$ vector of factor
loadings; ${\bff_t}$ is a $K\times1$ vector of common factors, and
$u_{it}$ is the idiosyncratic error component of $y_{it}$. Classical
factor analysis assumes that both $p$ and $K$ are fixed, while $T$ is
allowed to grow. However, in the recent decades, both economic and
financial applications have encountered very large data sets which
contain high-dimensional variables. For example, the World Bank has
data for about two hundred countries over forty\vadjust{\goodbreak} years; in portfolio
allocation, the number of stocks can be in thousands and be larger or
of the same order of the sample size. In modeling housing prices in
each zip code, the number of regions can be of order thousands, yet the
sample size can be 240 months or twenty years. The covariance matrix of
order several thousands is critical for understanding the co-movement
of housing prices indices over these zip codes.

Inferential theory of factor analysis relies on estimating $\Sig_u$,
the variance--covariance matrix of the error term, and $\Sig$, the
variance--covariance matrix of $\by_t=(y_{1t},\ldots,y_{pt})'$. In the
literature, $\Sig=\cov(\by_t)$ was traditionally estimated by the
sample covariance matrix of $\by_t$.
\[
\Sig_{\mathit{sam}}=\frac{1}{T-1}\sum_{t=1}^T(\by_t-\bar{\mathbf{y}})(\by
_t-
\bar{\mathbf{y}})',
\]
which was always assumed to be pointwise root-$T$ consistent. However,
the sample covariance matrix is an inappropriate estimator in
high-dimensional settings. For example, when $p$ is larger than $T$,
$\Sig_{\mathit{sam}}$ becomes singular while $\Sig$ is always strictly positive
definite. Even if $p<T$, \citet{FanFanLv08} showed that this
estimator has a very slow convergence rate under the Frobenius norm.
Realizing the limitation of the sample covariance estimator in
high-dimensional factor models, \citet{FanFanLv08} considered more
refined estimation of $\Sig$, by incorporating the common factor
structure. One of the key assumptions they made was the cross-sectional
independence among the idiosyncratic components, which results in a
diagonal matrix $\Sig_u=E\bu_t\bu_t'$. The cross-sectional
independence, however, is restrictive in many applications, as it rules
out the \textit{approximate factor structure} as in \citet{Cha83}.
In this paper, we relax this assumption, and investigate the impact of
the cross-sectional correlations of idiosyncratic noises on the
estimation of $\Sig$ and $\Sig_u$, when both $p$ and $T$ are allowed to
diverge. We show that the estimated covariance matrices are still
invertible with probability approaching one, even if $p>T$. In
particular, when estimating $\Sig^{-1}$ and $\Sig _u^{-1}$, we allow
$p$ to increase much faster than $T$, say, $p=O(\exp (T^{\alpha}))$,
for some $\alpha\in(0,1)$.

Sparsity is one of the commonly used assumptions in the estimation of
high-dimensional covariance matrices, which assumes that many entries
of the off-diagonal elements are zero, and the number of nonzero
off-diagonal entries is restricted to grow slowly. Imposing the
sparsity assumption directly on the covariance of $\by_t$, however, is
inappropriate for many applications of finance and economics. In this
paper we use the factor model and assume that $\Sig_u$ is sparse, and
estimate both $\Sig_u$ and $\Sig_u^{-1}$ using the thresholding method
[\citet{BicLev08N1}, \citet{CaiLiu11}] based on the estimated
residuals in the factor model. It is assumed that the factors ${\bff
_t}$ are observable, as in \citet{FamFre92}, \citet{FanFanLv08}, and many other empirical applications. We derive the
convergence rates of both estimated $\Sig$ and its inverse,
respectively,\vadjust{\goodbreak}
under various norms which are to be defined later. In addition, we
achieve better convergence rates than those in \citet{FanFanLv08}.

Various approaches have been proposed to estimate a large covariance
matrix: Bickel and Levina (\citeyear{BicLev08N1},
\citeyear{BicLev08N2}) constructed the estimators
based on regularization and thresholding, respectively.
\citet{RotLevZhu09}
considered thresholding the sample covariance matrix
with more general thresholding functions. \citet{LamFan09} proposed
penalized quasi-likelihood method to achieve both the consistency and
sparsistency of the estimation. More recently, \citet{CaiZho}
derived the minimax rate for sparse matrix estimation, and showed that
the thresholding estimator attains this optimal rate under the operator
norm. \citet{CaiLiu11} proposed a thresholding procedure which is
adaptive to the variability of individual entries and unveiled its
improved rate of convergence.

The rest of the paper is organized as follows. Section \ref{sec2}
provides the asymptotic theory for estimating the error covariance
matrix and its inverse. Section \ref{sec3} considers estimating the
covariance matrix of $\by _t$. Section \ref{sec4} extends the results
to the \textit{seemingly unrelated regression} model, a set of linear
equations with correlated error terms in which the covariates are
different across equations. Section \ref{sec5} reports the simulation
results. Finally, Section \ref{sec6} concludes with discussions. All proofs are
given in the \hyperref[app]{Appendix}. Throughout the paper, we use
$\lambda_{\min}(\bA)$ and $\lambda_{\max}(\bA)$ to denote the minimum
and maximum eigenvalues of a matrix $\bA$. We also denote by
$\|\bA\|_F$, $\|\bA\|$ and $\|\bA\|_{\Max}$ the Frobenius norm,
operator norm and elementwise norm of a matrix $\bA$, respectively,
defined, respectively, as $\|\bA\|_F=\tr^{1/2}(\bA'\bA)$, $\|\bA\|
=\lambda _{\max}^{1/2}(\bA'\bA)$ and $\|\bA\|_{\Max}=\max
_{i,j}|A_{ij}|$. Note that, when $\bA$ is a vector, both $\|\bA\|$ and~$\|\bA\|_F$ are equal to the Euclidean norm.

\section{Estimation of error covariance matrix}\label{sec2}
\subsection{Adaptive thresholding}
Consider the following approximate factor model, in which the
cross-sectional correlation among the idiosyncratic error components is allowed:
%
\begin{equation}\label{e21}
y_{it}={\mathbf b}_i'{\bff_t}+u_{it},
\end{equation}
where $i=1,\ldots,p$ and $t=1,\ldots,T$; ${\mathbf b}_i$ is a $K\times1$
vector of
factor loadings; ${\bff_t}$ is a $K\times1$ vector of observable
common factors, uncorrelated with $u_{it}$. Write
\[
\bB=({\mathbf b}_1,\ldots,{\mathbf b}_p)',\qquad
\by_t=(y_{1t},\ldots,y_{pt})';\qquad
\bu_t=(u_{1t},\ldots,u_{pt})',
\]
then model (\ref{e21}) can be written in a more compact form,
%
\begin{equation}
\by_t=\bB\bff_t+{\bu_t}
\end{equation}
with $E(\bu_t|\bff_t)=0$.

In practical applications, $p$ can be thought of as the number of
assets or stocks, or number of regions in spatial and temporal problems
such as home price\vadjust{\goodbreak} indices or sales of drugs, and in practice can be of
the same order as, or even larger than~$T$. For example, an asset
pricing model may contain hundreds of assets while the sample size on
daily returns is less than several hundreds. In the estimation of the
optimal portfolio allocation, it was observed by
\citet{FanFanLv08} that the effect of large $p$ on the convergence
rate can be quite severe. In contrast, the number of common factors,
$K$, can be much smaller. For example, the rank theory of consumer
demand systems implies no more than three factors [e.g.,
\citet{Gor81} and \citet{Lew91}].

The error covariance matrix
\[
\Sig_u=\cov({\bu_t}),
\]
itself is of interest for the inferential theory of factor models. For
example, the asymptotic covariance of the least square estimator of
$\bB
$ depends on~$\Sig_u^{-1}$, and in simulating home price indices over a
certain time horizon for mortgage based securities, a good estimate of
$\Sig_u$ is needed. When $p$ is close to or larger than~$T$, estimating
$\Sig_u$ is very challenging. Therefore, following the literature of
high-dimensional covariance matrix estimation, we assume it is sparse,
that is, many of its off-diagonal entries are zeros. Specifically, let
$\Sig_u=(\sigma_{ij})_{p\times p}$. Define
%
\begin{equation}\label{e23}
m_T=\max_{i\leq p}\sum_{ j\leq p}I(\sigma_{ij}\neq0).
\end{equation}
The sparsity assumption puts an upper bound restriction on $m_T$.
Specifically, we assume
%
\begin{equation}
m_T^2=o\biggl(\frac{T}{K^2\log p}\biggr).
\end{equation}
In this formulation, we even allow the number of factors $K$ to be
large, possibly growing with $T$.

A more general treatment [e.g., \citet{BicLev08N1} and
\citet{CaiLiu11}] is to assume that the $l_q$ norm of the row
vectors of $\Sig _u$ are uniformly bounded across rows by a slowly
growing sequence, for some $q\in[0,1)$. In contrast, the assumption we
make in this paper, that is, $q=0$, has clearer economic
interpretation. For example, the firm returns can be modeled by the
factor model, where $u_{it}$ represents a firm's individual shock at
time~$t$. Driven by the industry-specific components, these shocks are
correlated among the firms in the same industry, but can be assumed to
be uncorrelated across industries, since the industry-specific
components are not pervasive for the whole economy
[\citet{ConKor93}].

We estimate $\Sig_u$ using the thresholding technique introduced and
studied by \citet{BicLev08N1}, \citet{RotLevZhu09},
\citet{CaiLiu11}, etc., which is summarized as follows: Suppose we
observe data $(\bX_1,\ldots,\bX_T)$ of a $p\times1$ vector $\bX$, which
follows\vadjust{\goodbreak} a multivariate Gaussian distribution $N(0,\Sig_X)$. The sample
covariance matrix of $\bX$ is thus given by
\[
\bS_X=\frac{1}{T}\sum_{i=1}^T(\bX_i-\bar{\bX})(\bX_i-\bar{\bX
})'=(s_{ij})_{p\times p}.
\]
Define\vspace*{2pt} the thresholding operator by $\mathcal{T}_t(\bM
)=(M_{ij}I(|M_{ij}|\geq t))$ for any symmetric matrix $\bM$. Then
$\mathcal{T}_t$ preserves the symmetry of $\bM$. Let $\hSig
{}^{\mathcal
{T}}_X=\mathcal{T}_{\omega_T}(\bS_X)$, where $\omega_T=O(\sqrt{\log
p/T})$. \citet{BicLev08N1} then showed that
\[
\|\hSig{}^{\mathcal{T}}_X-\Sig_X\|=O_p(\omega_Tm_T).
\]

In the factor models, however, we do not observe the error term
directly. Hence when estimating the error covariance matrix of a factor
model, we need to construct a sample covariance matrix based on the
residuals $\hat{u}_{it}$ before thresholding. The residuals are
obtained using the plug-in method, by estimating the factor loadings
first. Let $\widehat{\mathbf b}_i$ be the ordinary least square
(OLS) estimator of ${\mathbf b}_i$, and
\[
\hat{u}_{it}=y_{it}-\widehat{\mathbf b}_i'{\bff_t}.
\]
Denote by $\hu_t =(\hat{u}_{1t},\ldots,\hat{u}_{pt})'$. We then construct
the residual covariance matrix as
\[
\hSig_u=\frac{1}{T}\sum_{t=1}^T\hu_t \hu_t'=(\widehat\sigma_{ij}).
\]
Note that the thresholding value $\omega_T=O(\sqrt{\log p/T})$ in
\citet{BicLev08N1} is in fact obtained from the rate of
convergence of $\max_{ij}|s_{ij}-\Sig_{X,ij}|$. This rate changes when
$s_{ij}$ is replaced with the residual $\hat{u}_{ij}$, which will be
slower if the number of common factors $K$ increases with $T$.
Therefore, the thresholding value $\omega_T$ used in this paper is
adjusted to account for the effect of the estimation of the residuals.

\subsection{Asymptotic properties of the thresholding estimator}

Bickel and Levina (\citeyear{BicLev08N1}) used a universal constant as the thresholding
value. As pointed out by \citet{RotLevZhu09} and \citet{CaiLiu11}, when the variances of the entries of the sample covariance
matrix vary over a wide range, it is more desirable to use thresholds
that capture the variability of individual estimation. For this
purpose, in this paper, we apply the adaptive thresholding estimator
[\citet{CaiLiu11}] to estimate the error covariance matrix, which is
given by
%
\begin{eqnarray}\label{e25}
\hSig_{u}^{\mathcal{T}}&=&(\widehat\sigma_{ij}^{\mathcal{T}}),\qquad
\widehat\sigma_{ij}^{\mathcal{T}}=\widehat\sigma_{ij}I\bigl(|\widehat
\sigma_{ij}|\geq
\sqrt{\hat{\theta}_{ij}}\omega_T\bigr),\nonumber\\[-9.5pt]\\[-9.5pt]
\hat{\theta}_{ij}&=&\frac{1}{T}\sum_{t=1}^T(\hat{u}_{it}
\hat{u}_{jt}-\widehat\sigma_{ij})^2\nonumber
\end{eqnarray}
for some $\omega_T$ to be specified later.\vadjust{\goodbreak}

We impose the following assumptions:
%
\begin{assum}\label{a21} (i)
$\{\bu_t\}_{t\geq1}$ is stationary and ergodic such that each~$\bu_t$
has zero mean vector and covariance matrix $\Sig_u$. In addition, the
strong mixing condition in Assumption \ref{a32} holds.

\mbox{}\hphantom{i}(ii) There exist constants $c_1, c_2>0$ such that $c_1<\lambda_{\min
}(\Sig_u)\leq\lambda_{\max}(\Sig_u)< c_2$, and $c_1<\var
(u_{it}u_{jt})<c_2$ for all $i\leq p$, $j\leq p$.

(iii) There exist $r_1>0$ and $b_1>0$, such that for any $s>0$ and
$i\leq p$,
%
\begin{equation}\label{e251}
P(|u_{it}|>s)\leq\exp\bigl(-(s/b_1)^{r_1}\bigr).
\end{equation}
\end{assum}

Condition (i) allows the idiosyncratic components to be weakly
dependent. We will formally present the strong mixing condition in the
next section. In order for the main results in this section to hold, it
suffices to impose the strong mixing condition marginally on $\bu_t$
only. Roughly speaking, we require the mixing coefficient
\[
\alpha(T)={\sup_{A\in\mathcal{F}_{-\infty}^0, B\in\mathcal
{F}_T^{\infty
}}}|P(A)P(B)-P(A\cap B)|
\]
to decrease exponentially fast as $T\rightarrow\infty$, where
$(\mathcal
{F}_{-\infty}^0, \mathcal{F}_T^{\infty})$ are the $\sigma$-algebras
generated by $\{\bu_t\}_{t=-\infty}^0$ and $\{\bu_t\}_{t=T}^{\infty}$,
respectively.

Condition (ii) requires the nonsingularity of $\Sig_u$. Note that
\citet{CaiLiu11} allowed $\max_j\sigma_{jj}$ to diverse when
direct observations are available. Condition~(ii), however, requires
that $\sigma_{jj}$ should be uniformly bounded. In factor models, a
uniform upper bound on the variance of $u_{it}$ is needed when we
estimate the covariance matrix of $\by_t$ later. This assumption is
satisfied by most of the applications of factor models. Condition (iii)
requires the distributions of $(u_{1t},\ldots, u_{pt})$ to have
exponential-type tails,\vspace*{1pt} which allows us to apply the large deviation
theory to $\frac {1}{T}\sum _{t=1}^Tu_{it}u_{jt}-\sigma_{ij}$.
%
\begin{assum}\label{a22} There exist positive sequences $\kappa_1(p,
T)=o(1)$, $\kappa_2(p,\allowbreak T)=o(1)$ and $a_T=o(1)$, and a constant $M>0$,
such that for all $C>M$,
\begin{eqnarray*}
P\Biggl(\max_{i\leq p}\frac{1}{T}\sum
_{t=1}^T|u_{it}-\hat{u}_{it}|^2>Ca_T^2\Biggr)&\leq&
O(\kappa_1(p,T)),\\
P\Bigl(\max_{i\leq p, t\leq T}|u_{it}-\hat{u}_{it}|>C\Bigr)&\leq&
O(\kappa_2(p,T)).
\end{eqnarray*}
\end{assum}

This assumption allows us to apply thresholding to the estimated error
covariance matrix when direct observations are not available, without
introducing too much extra estimation error. Note that it permits a
general case when the original ``data'' is contaminated, including any
type of estimate of the data when direct\vadjust{\goodbreak} observations are not
available, as well as the case when data is subject to measurement of
errors. We will show in the next section that in a linear factor model
when $\{u_{it}\}_{i\leq p, t\leq T}$ are estimated using the OLS
estimator, the rate of convergence $a_T^2=(K^2\log p)/T$.

The following\vspace*{1pt} theorem establishes the asymptotic properties of the
thresholding estimator $\hSig{}^{\mathcal{T}}_u$, based on observations
with estimation errors. Let $\gamma^{-1}=3r_1^{-1}+r_2^{-1}$, where
$r_1$ and $r_2$ are defined in Assumptions \ref{a21}, \ref{a32}, respectively.
%
\begin{theorem}\label{t21} Suppose\vspace*{1pt} $\gamma<1$ and $(\log p)^{6/\gamma
-1}=o(T)$. Then under Assumptions \ref{a21} and \ref{a22}, there exist
$C_1>0$ and $C_2>0$ such that for $\hSig{}^{\mathcal{T}}_u$ defined in
(\ref{e25}) with
\[
\omega_T=C_1\Biggl(\sqrt{\frac{\log p}{T}}+a_T\Biggr),
\]
we have
%
\begin{equation}\label{e26}
P(\|\hSig{}^{\mathcal{T}}_u-\Sig_u\|\leq C_2\omega_Tm_T
)\geq
1-O\biggl(\frac{1}{p^2}+\kappa_1(p, T)+\kappa_2(p, T)\biggr).
\end{equation}
In addition, if $\omega_Tm_T=o(1)$, then with probability at least $
1-O(\frac{1}{p^2}+\kappa_1(p, T)+\kappa_2(p, T))$,
\[
\lambda_{\min}(\hSig{}^{\mathcal{T}}_u)\geq0.5\lambda_{\min}(\Sig_u)
\]
and
\[
\|(\hSig{}^{\mathcal{T}}_u)^{-1}-\Sig_u^{-1}\|\leq C_2\omega_Tm_T.
\]
\end{theorem}

Note that we derive result (\ref{e26}) without assuming the sparsity on
$\Sig_u$, that is, no restriction is imposed on $m_T$. When $\omega
_Tm_T\neq o(1)$, (\ref{e26}) still holds, but $\|\hSig{}^{\mathcal{T}}_u
-\Sig_u\|$ does not converge to zero in probability. On the other
hand, the condition $\omega_Tm_T=o(1)$ is required to preserve the
nonsingularity of $\hSig{}^{\mathcal{T}}_u$ asymptotically and to
consistently estimate $\Sig_u^{-1}$.

The rate of convergence also depends on the averaged estimation error
of the residual terms. We will see in the next section that when the
number of common factors $K$ increases slowly, the convergence rate in
Theorem \ref{t21} is close to the minimax optimal rate as in
\citet{CaiZho}.

\section{Estimation of covariance matrix using factors}\label{sec3}

We now investigate the estimation of the covariance matrix $\Sig$ in
the approximate factor model
\[
\by_t=\bB\bff_t+{\bu_t},
\]
where $\Sig=\cov(\by_t)$. This covariance matrix is particularly of
interest in many applications of factor models as well as corresponding
inferential theories. When estimating a large dimensional covariance
matrix, sparsity and banding are two commonly used assumptions for
regularization [e.g., Bickel and Levina
(\citeyear{BicLev08N1}, \citeyear{BicLev08N2})]. In most of the
applications in finance and economics, however, these two assumptions
are inappropriate for $\Sig$. For instance, the US housing prices in
the county level are generally associated with a few national indices,
and there is no natural ordering among the counties. Hence neither the
sparsity nor the banding is realistic for such a~problem. On the other
hand, it is natural to assume $\Sig_u$ sparse, after controlling the
common factors. Therefore, our approach combines the merits of both the
sparsity and factor structures.

Note that
\[
\Sig={\bB}\cov({\bff_t}){\bB}'+\Sig_u.
\]
By the Sherman--Morrison--Woodbury formula,
\[
\Sig^{-1}=\Sig_u^{-1}-\Sig_u^{-1}{\bB}[\cov({\bff_t})^{-1}+{\bB
}'\Sig
_u^{-1}{\bB}]^{-1}{\bB}'\Sig_u^{-1}.
\]

When the factors are observable, one can estimate $\bB$ by the least
squares method, $\hB=({\widehat{\mathbf b}}_1,\ldots,{\widehat
{\mathbf b}}_p)'$, where
\[
\widehat{\mathbf b}_i=\arg\min_{{\mathbf b}_i}\frac
{1}{Tp}\sum_{t=1}^T\sum_{i=1}^p(
y_{it}-{\mathbf b}_i'{\bff_t})^2.
\]
The covariance matrix $\cov({\bff_t})$ can be estimated by the sample
covariance matrix
\[
\hcov({\bff_t})=T^{-1}\bX\bX'-T^{-2}\bX\bone\bone'\bX',
\]
where $\bX=(\bff_1,\ldots,\bff_T)$, and $\bone$ is a $T$-dimensional
column vector of ones. Therefore, by employing the thresholding
estimator $\hSig{}^{\mathcal{T}}_u$ in (\ref{e25}), we obtain
substitution estimators
%
\begin{equation} \label{ee31}
\hSig{}^{\mathcal{T}}=\hB\hcov({\bff_t})\hB'+\hSig{}^{\mathcal{T}}_u
\end{equation}
and
%
\begin{equation} \label{e32}\qquad
(\hSig{}^{\mathcal{T}})^{-1}=(\hSig{}^{\mathcal{T}}_u)^{-1}-(
\hSig{}^{\mathcal{T}}_u
)^{-1}\hB[\hcov({\bff_t})^{-1}+\hB'(
\hSig{}^{\mathcal{T}}_u)^{-1}\hB]^{-1}\hB'(\hSig{}^{\mathcal{T}}_u)^{-1}.
\end{equation}

In practice, one may apply a common thresholding $\lambda$ to the
correlation matrix of $\hSig_u$, and then use the substitution
estimator similar to (\ref{ee31}). When $\lambda=0$ (no thresholding),
the resulting estimator is the sample covariance, whereas when \mbox{$\lambda
=1$} (all off-diagonals are thresholded), the resulting estimator is an
estimator based on the strict factor model [\citet{FanFanLv08}].
Thus we have created a path (indexed by $\lambda$) which connects the
nonparametric estimate of covariance matrix to the parametric estimate.

The following assumptions are made:
%
\begin{assum} \label{a31}(i) $\{\bff_t\}_{t\geq1}$ is stationary and
ergodic.

(ii) $\{{\bu}_t\}_{t\geq1}$ and $\{\bff_t\}_{t\geq1}$ are
independent.\vadjust{\goodbreak}
\end{assum}

In addition\vspace*{1pt} to the conditions above, we introduce the strong mixing
conditions to conduct asymptotic analysis of the least square
estimates. Let~$\mathcal{F}_{-\infty}^0$ and $\mathcal
{F}_{T}^{\infty}$
denote the $\sigma$-algebras generated by $\{(\bff_t,\bu_t)\dvtx -\infty
\leq t\leq0\}$ and $\{(\bff_t,\bu_t)\dvtx T\leq t\leq\infty\}$,
respectively. In addition, define the mixing coefficient
\[
\alpha(T)={\sup_{A\in\mathcal{F}_{-\infty}^0, B\in\mathcal
{F}_{T}^{\infty
}}}|P(A)P(B)-P(AB)|.
\]
The following strong mixing assumption enables us to apply the
Bernstein's inequality in the technical proofs.
%
\begin{assum} \label{a32} There exist positive constants $r_2$ and $C$
such that for all $t\in\mathbb{Z}^+$,
\[
\alpha(t)\leq\exp(-Ct^{r_2}).
\]
\end{assum}

In addition, we impose the following regularity conditions:
%
\begin{assum}\label{a33} (i) There exists a constant $M>0$ such that
for all~$i,j$ and~$t$, $Ey_{it}^2<M$, $Ef_{it}^2<M$ and $|b_{ij}|<M$.

(ii) There exists a constant $r_3>0$ with $3r_3^{-1}+r_2^{-1}>1$, and
$b_2>0$ such that for any $s>0$ and $i\leq K$,
%
\begin{equation} \label{e333}
P(|f_{it}|>s)\leq\exp\bigl(-(s/b_2)^{r_3}\bigr).
\end{equation}
\end{assum}

Condition (ii) allows us to apply the Bernstein-type inequality for the
weakly dependent data.
%
\begin{assum}\label{a34} There exists a constant $C>0$ such that\break
\mbox{$\lambda_{\min}(\cov(\bff_t))>C$}.
\end{assum}

Assumptions \ref{a34} and \ref{a21} ensure that both $\lambda_{\min
}(\cov(\bff_t))$ and $\lambda_{\min}(\Sig)$ are bounded away\vspace*{1pt} from zero,
which is needed to derive the convergence rate of $\|(\hSig{}^{\mathcal
{T}})^{-1}-\Sig^{-1}\|$ below.

The following lemma verifies Assumption \ref{a22}, which derives the
rate of convergence of the OLS estimator as well as the estimated residuals.

Let $\gamma_2^{-1}=1.5r_1^{-1}+1.5r_3^{-1}+r_2^{-1}$.
%
\begin{lem}\label{l31} Suppose $K=o(p)$, $K^4(\log p)^2=o(T)$ and
$(\log p)^{2/\gamma_2-1}=o(T)$. Then under the assumptions of Theorem
\ref{t21} and Assumptions \mbox{\ref{a31}--\ref{a34}}, there exists $C>0$,
such that:

\begin{longlist}
\item
\[
P\Biggl(\max_{i\leq p}\|\widehat{\mathbf b}_i-\mathbf{b}_i\|
>C\sqrt{\frac{K\log p}{T}}
\Biggr)=O\biggl(\frac{1}{p^2}+\frac{1}{T^2}\biggr);
\]

\item
\[
P\Biggl(\max_{i\leq p}\frac{1}{T}\sum_{t=1}^T|u_{it}-\hat
{u}_{it}|^2>\frac{CK^2\log p}{T}\Biggr)=O\biggl(\frac{1}{p^2}+\frac
{1}{T^2}\biggr);
\]

\item
\[
P\Biggl(\max_{i\leq p,t\leq T}|u_{it}-\hat{u}_{it}|>CK(\log
T)^{1/r_3}\sqrt{\frac{\log p}{T}}\Biggr)=O\biggl(\frac
{1}{p^2}+\frac
{1}{T^2}\biggr).
\]
\end{longlist}
\end{lem}

By Lemma \ref{l31} and Assumption \ref{a22}, $a_T=K\sqrt{(\log p)/T}$
and $\kappa_1(p, T)=\kappa_2(p, T)=p^{-2}+T^{-2}$. Therefore in the
linear approximate factor model, the thresholding parameter $\omega_T$
defined in Theorem \ref{t21} is simplified to the following: for some
positive constant $C_1'$,
%
\begin{equation} \label{e33}
\omega_T=C_1'K\sqrt{\frac{\log p}{T}}.
\end{equation}

Now we can apply Theorem \ref{t21} to obtain the following theorem:
%
\begin{theorem} \label{t31} Under the assumptions of Lemma \ref{l31},
there exist $C_1'>0$ and $C_2'>0$ such that the adaptive thresholding
estimator defined in (\ref{e25}) with $\omega_T^2=C_1'\frac{K^2\log
p}{T}$ satisfies:

\begin{longlist}
\item
\[
P\Biggl(\|\hSig{}^{\mathcal{T}}_u-\Sig_u\|\leq C_2'm_TK\sqrt{\frac
{\log
p}{T}}\Biggr)=1-O\biggl(\frac{1}{p^2}+\frac{1}{T^2}\biggr).
\]
\item If $m_TK\sqrt{\frac{\log p}{T}}=o(1)$, then with probability at
least $1-O(\frac{1}{p^2}+\frac{1}{T^2})$,
\[
\lambda_{\min}(\hSig{}^{\mathcal{T}}_u)\geq0.5\lambda_{\min}(\Sig_u)
\]
and
\[
\|(\hSig{}^{\mathcal{T}}_u)^{-1}-\Sig_u^{-1}\|\leq C_2'm_TK\sqrt
{\frac
{\log p}{T}}.
\]
\end{longlist}
\end{theorem}
%
\begin{remark}
We briefly comment on the terms in the convergence rate above.
\begin{longlist}[(2)]
\item[(1)] The term $K$ appears as an effect of using the estimated
residuals to construct the thresholding covariance estimator, which is
typically small compared to $p$ and $T$ in many applications. For
instance, the famous Fama--French three-factor model shows that $K=3$
factors are adequate for the US equity market. In an empirical study on
asset returns, \citet{BaiNg02} used the monthly data which contains
the returns of 4883 stocks for sixty months. For their data set,
$T=60$, $p=4883$. \citet{BaiNg02} determined $K=2$ common factors.
\item[(2)] As in \citet{BicLev08N1} and \citet{CaiLiu11},
$m_T$, the maximum number of nonzero components across the rows of
$\Sig
_u$, also plays a role in the convergence rate. Note that when $K$ is
bounded, the convergence rate reduces to $O_p(m_T\sqrt{(\log
p)/T})$, the same as the minimax rate derived by \citet{CaiZho}.
\end{longlist}
\end{remark}

One of our main objectives is to estimate $\Sig$, which is the
$p\times
p$ dimensional covarinace matrix of $\by_t$, assumed to be time
invariant. We can achieve a better accuracy in estimating both $\Sig$
and $\Sig^{-1}$ by incorporating the factor structure than using the
sample covariance matrix, as shown by \citet{FanFanLv08} in the strict
factor model case. When the cross-sectional correlations among the
idiosyncratic components $(u_{1t},\ldots,u_{pt})$ are in presence, we can
still take advantage of the factor structure. This is particularly
essential when direct sparsity assumption on $\Sig$ is inappropriate.
%
\begin{assum}\label{a35} $\|p^{-1}{\bB}'{\bB}-\bOmega\|=o(1)$ for some
$K\times K$ symmetric positive definite matrix $\bOmega$ such that
$\lambda_{\min}(\bOmega)$ is bounded away from zero.
\end{assum}

Assumption \ref{a35} requires that the factors should be pervasive,
that is, impact every individual time series [\citet{Har09}]. It was
imposed by \citet{FanFanLv08} only when they tried to establish the
asymptotic normality of the covariance estimator. However, it turns out
to be also
helpful to obtain a good upper bound of $\|(\hSig{}^{\mathcal
{T}})^{-1}-\Sig^{-1}\|$, as it ensures that $\lambda_{\max}(({\bB
}'\Sig
^{-1}{\bB})^{-1})=O(p^{-1})$.\vspace*{1pt}

\citet{FanFanLv08} obtained an upper bound of $\|\hSig{}^{\mathcal
{T}}-\Sig
\|_F$ under the Frobenius norm when $\Sig_u$ is diagonal, that is,
there was no cross-sectional correlation among the idiosyncratic
errors. In order for their upper bound to decrease to zero, $p^2<T$ is
required. Even with this restrictive assumption, they showed that the
convergence rate is the same as the usual sample covariance matrix of
$\by_t$, though the latter does not take the factor structure into
account. Alternatively, they considered the entropy loss norm, proposed
by \citet{JamSte61},
\[
\|\hSig{}^{\mathcal{T}}-\Sig\|_{\Sigma}=\bigl(p^{-1}\tr[(\hSig
{}^{\mathcal
{T}}\Sig^{-1}-I)^2]\bigr)^{1/2}=p^{-1/2}\|\Sig^{-1/2}(\hSig
{}^{\mathcal
{T}}-\Sig)\Sig^{-1/2}\|_F.
\]
Here the factor $p^{-1/2}$ is used for normalization, such that $\|\Sig
\|_{\Sigma}=1$. Under this norm, \citet{FanFanLv08} showed that the
substitution estimator has a better convergence rate than the usual
sample covariance matrix. Note that the normalization factor $p^{-1/2}$
in the definition results in an averaged estimation error, which also
cancels out the diverging dimensionality introduced by $p$. In
addition, for any two $p\times p$ matrices $\bA_1$ and $\bA_2$,
\begin{eqnarray*}
\|\bA_1-\bA_2\|_{\Sigma}&=&p^{-1/2}\|\Sig^{-1/2}(\bA_1-\bA_2)\Sig
^{-1/2}\|_F\\
&\leq&\|\Sig^{-1/2}(\bA_1-\bA_2)\Sig^{-1/2}\|\\
&\leq& \|\bA_1-\bA_2\|\cdot\lambda_{\max}(\Sig^{-1}).
\end{eqnarray*}

Combining with the estimated low-rank matrix $\bB\cov(\bff_t)\bB'$,
Theorem \ref{t31} implies the main theorem in this section:
%
\begin{theorem} \label{t32} Suppose $\log T=o(p)$. Under the
assumptions of Theorem~\ref{t31} and Assumption \ref{a35}, we have:

\begin{longlist}
\item
\begin{eqnarray*}
&&
P\biggl(\|\hSig{}^{\mathcal{T}}-\Sig\|_{\Sigma}^{2}\leq\frac
{CpK^2(\log
p)^2}{T^2}+\frac{Cm_T^2K^2\log
p}{T}\biggr)\\
&&\qquad=1-O\biggl(\frac{1}{p^2}+\frac{1}{T^2}\biggr),\\
&& P\biggl(\|\hSig{}^{\mathcal{T}}-\Sig\|_{\Max}^2\leq\frac
{CK^2\log
p+CK^4\log T}{T}\biggr)\\
&&\qquad=1-O\biggl(\frac{1}{p^2}+\frac{1}{T^2}\biggr).
\end{eqnarray*}
\item If $m_TK\sqrt{\frac{\log p}{T}}=o(1)$, with probability at least
$1-O(\frac{1}{p^2}+\frac{1}{T^2})$,
\[
\lambda_{\min}(\hSig{}^{\mathcal{T}})\geq0.5\lambda_{\min}(\Sig_u)
\]
and
\[
\|(\hSig{}^{\mathcal{T}})^{-1}-\Sig^{-1}\|\leq Cm_TK\sqrt{\frac{\log p}{T}}.
\]
\end{longlist}
\end{theorem}

Note that we have derived a better convergence rate of $(\hSig
{}^{\mathcal
{T}})^{-1}$ than that in \citet{FanFanLv08}. When the operator norm is
considered, $p$ is allowed to grow exponentially fast in $T$ in order
for $(\hSig{}^{\mathcal{T}})^{-1}$ to be consistent.

We have also derived the maximum elementwise estimation $\|\hSig
{}^{\mathcal{T}}-\Sig\|_{\Max}$. This quantity appears in risk
assessment as in \citet{FanZhaYu08}. For any portfolio with
allocation vector $\bw$, the true portfolio variance and the estimated
one are given by $\bw'\Sig\bw$ and $\bw'\hSig{}^{\mathcal{T}}\bw$,
respectively. The estimation error is bounded by
\[
|\bw'\hSig{}^{\mathcal{T}}\bw-\bw'\Sig\bw|\leq\|\hSig{}^{\mathcal
{T}}-\Sig
\|_{\Max}\|\bw\|_1^2,
\]
where $\|\bw\|_1$, the $l_1$ norm of $\bw$, is the gross exposure of
the portfolio.

\section{Extension: Seemingly unrelated regression}\label{sec4}

A \textit{seemingly unrelated regression} model [\citet{KmeGil70}] is a set of linear equations in which the disturbances are
correlated across equations. Specifically, we have
%
\begin{equation} \label{e41}
y_{it}=\mathbf{b}_i'\bff_{it}+u_{it},\qquad i\leq p, t\leq T,
\end{equation}
where $\mathbf{b}_i$ and $\bff_{it}$ are both $K_i\times1$
vectors. The $p$
linear equations (\ref{e41}) are related because their error terms
$u_{it}$ are correlated; that is, the covariance matrix
\[
\Sig_u=(Eu_{it}u_{jt})_{p\times p}
\]
is not diagonal.

Model (\ref{e41}) allows each variable $y_{it}$ to have its own
factors. This is important for many applications. In financial
applications, the returns of individual stock depend on common market
factors and sector-specific factors. In housing price index modeling,
housing price appreciations depend on both national factors and local
economy. When $\bff_{it}=\bff_{t}$ for each $i\leq p$, model (\ref
{e41}) reduces to the approximate factor model (1.1) with common
factors $\bff_t$.

Under mild conditions, running OLS on each equation produces unbiased
and consistent estimator of $\mathbf{b}_i$ separately. However,
since OLS does
not take into account the cross-sectional correlation among the noises,
it is not efficient. Instead, statisticians obtain the best linear
unbiased estimator (BLUE) via generalized least square (GLS). Write
\begin{eqnarray*}
\by_i&=&(y_{i1},\ldots,y_{iT})', {T\times1}, \qquad\bX_i=(\bff
_{i1},\ldots,\bff_{iT})', {T\times K_i},\qquad i\leq p,
\\
\by&=&\pmatrix{
\by_1 \cr
\vdots\cr
\by_p },\qquad
\bX=\pmatrix{\bX_1 & 0 & 0 \cr
0 & \ddots& 0 \cr
0 & 0 & \bX_p},\qquad
\bB=\pmatrix{\mathbf{b}_1 \cr
\vdots\cr
\mathbf{b}_p}.
\end{eqnarray*}
The GLS estimator of $\bB$ is given by \citet{Zel62}.
%
\begin{equation} \label{e42}
\hB_{\mathrm{GLS}}=[\bX'(\hSig{}^{-1}_u\otimes I_T)^{-1}\bX]^{-1}[\bX'(
\hSig{}^{-1}_u
\otimes I_T)^{-1}\by],
\end{equation}
where $I_T$ denotes a $T\times T$ identity matrix, $\otimes$ represents
the Kronecker product operation and $\hSig_u$ is a consistent estimator
of $\Sig_u$.

In classical seemingly unrelated regression in which $p$ does not grow
with~$T$, $\Sig_u$ is estimated by a two-stage procedure
[\citet{KmeGil70}]: In the first stage, estimate $\bB$ via OLS,
and obtain residuals
%
\begin{equation} \label{e43}
\hat{u}_{it}=y_{it}-\widehat{\mathbf b}_{i}'\bff_{it}.
\end{equation}
In the second stage, estimate $\Sig_u$ by
%
\begin{equation}
\hSig_{u}=(\widehat\sigma_{ij})=\Biggl(\frac{1}{T}\sum_{t=1}^T\hat
{u}_{it}\hat
{u}_{jt}\Biggr)_{p\times p}.
\end{equation}
In high dimensional, seemingly unrelated regression in which $p>T$,
however, $\hSig_u$~is not invertible, and hence the GLS estimator
(\ref
{e42}) is infeasible.\vadjust{\goodbreak}

By the sparsity assumption of $\Sig_u$, we can deal with this
singularity problem by using the adaptive thresholding estimator, and
produce a consistent nonsingular estimator of $\Sig_u$,
%
\begin{equation} \label{e44}
\hSig{}^{\mathcal{T}}_u=\bigl(\widehat\sigma_{ij}I\bigl(|\widehat\sigma
_{ij}|>\sqrt{\hat{\theta
}_{ij}}\omega_T\bigr)\bigr),\qquad\hat{\theta}_{ij}=\frac
{1}{T}\sum
_{t=1}^T(\hat{u}_{it}\hat{u}_{jt}-\widehat\sigma_{ij})^2.
\end{equation}

To pursue this goal, we impose the following assumptions:
%
\begin{assum} \label{a41} For each $i\leq p$:

\begin{longlist}
\item
$\{\bff_{it}\}_{t\geq1}$ is stationary and ergodic.
\item
$\{\bu_t\}_{t\geq1}$ and $\{\bff_{it}\}_{t\geq1}$ are independent.
\end{longlist}
\end{assum}
%
\begin{assum} There exists positive constants $C$ and $r_2$ such that
for each $i\leq p$, the strong mixing condition
\[
\alpha(t)\leq\exp(-Ct^{r_2})
\]
is satisfied by $(\bff_{it}, \bu_t)$.
\end{assum}
%
\begin{assum} \label{a43} There exist constants $M$ and $C>0$ such that
for all $i\leq p, j\leq K_i, t\leq T$:

\begin{longlist}
\item
$Ey_{it}^2<M$, $|b_{ij}|<M$ and $Ef_{it, j}^2<M$.

\item
$\min_{i\leq p}\lambda_{\min}(\cov(\bff_{it}))>C$.
\end{longlist}
\end{assum}
%
\begin{assum} \label{a44} There exists a constant $r_4>0$ with
$3r_4^{-1}+r_2^{-1}>1$, and $b_3>0$ such that for any $s>0$ and $i,j$,
\[
P(|f_{it,j}|>s)\leq\exp\bigl(-(s/b_3)^{r_4}\bigr).
\]
\end{assum}

These assumptions are similar to those made in Section \ref{sec3}, except that
here they are imposed on the sector-specific factors.
The main theorem in this section is a direct application of Theorem
\ref
{t21}, which shows that the adaptive thresholding produces a consistent
nonsingular estimator of $\hSig_u$.

\begin{theorem}\label{t41}
Let $K=\max_{i\leq p}K_i$ and $\gamma
_3^{-1}=1.5r_1^{-1}+1.5r_4^{-1}+r_2^{-1}$; suppose $K=o(p)$, $K^4(\log
p)^2=o(T)$ and $(\log p)^{2/\gamma_3-1}=o(T)$. Under Assumptions \ref
{a21}, \ref{a41}--\ref{a44}, there exist constants $C_1>0$ and $C_2>0$
such that the adaptive thresholding estimator defined in (\ref{e44})
with $\omega_T^2=C_1\frac{K^2\log p}{T}$
satisfies:

\begin{longlist}
\item
\[
P\Biggl(\|\hSig{}^{\mathcal{T}}_u-\Sig_u\|\leq C_2m_TK\sqrt{\frac
{\log
p}{T}}\Biggr)=1-O\biggl(\frac{1}{p^2}+\frac{1}{T^2}\biggr).
\]
\item If $m_TK\sqrt{\frac{\log p}{T}}=o(1)$, then with probability at
least $1-O(\frac{1}{p^2}+\frac{1}{T^2})$,
\[
\lambda_{\min}(\hSig{}^{\mathcal{T}}_u)\geq0.5\lambda_{\min}(\Sig_u)
\]
and
\[
\|(\hSig{}^{\mathcal{T}}_u)^{-1}-\Sig_u^{-1}\|\leq C_2m_TK\sqrt{\frac
{\log p}{T}}.
\]
\end{longlist}
\end{theorem}

Therefore, in the case when $p>T$, Theorem \ref{t41} enables us to
efficiently estimate $\bB$ via feasible GLS.
\[
\hB_{\mathrm{GLS}}^{\mathcal{T}}=\bigl[\bX'\bigl((\hSig{}^{\mathcal{T}}_u)^{-1}\otimes
I_T\bigr)^{-1}\bX\bigr]^{-1}\bigl[\bX'\bigl((\hSig{}^{\mathcal{T}}_u)^{-1}\otimes
I_T\bigr)^{-1}\by\bigr].
\]

\section{Monte Carlo experiments}\label{sec5}

In this section, we use simulation to demonstrate the rates of
convergence of the estimators $\hSig{}^{\mathcal{T}}$ and $(\hSig
{}^{\mathcal{T}})^{-1}$ that we have obtained so far. The simulation
model is a modified version of the Fama--French three-factor model
described in \citet{FanFanLv08}. We fix the number of factors, $K=3$,
and the length of time, $T=500$, and let the dimensionality $p$
gradually increase.

The Fama--French three-factor model [\citet{FamFre92}] is given~by
\[
y _{it}=b_{i1}f_{1t}+b_{i2}f_{2t}+b_{i3}f_{3t}+u_{it},
\]
which models the excess return (real rate of return minus risk-free
rate) of the $i$th stock of a portfolio, $y_{it}$, with respect
to 3 factors. The first factor is the excess return of the whole stock
market, and the weighted excess return on all NASDAQ, AMEX and NYSE
stocks is a commonly used proxy. It extends the capital assets pricing
model (CAPM) by adding two new factors---SMB (``small minus big'' cap)
and HML (``high minus low'' book/price). These two were added to the
model after the observation that two types of stocks---small caps, and
high book value to price ratio---tend to outperform the stock market as
a whole.

We separate this section into three parts, calibration, simulation and
results. Similarly to Section 5 of \citet{FanFanLv08}, in the
calibration part we want to calculate realistic multivariate
distributions from which we can generate the factor loadings ${\bB}$,
idiosyncratic noises $\{\bu_t\}_{t=1}^T$ and the observable factors $\{
\bff_t\}_{t=1}^T$. The data was obtained from the data library of
Kenneth French's website.

\subsection{Calibration}
To estimate the parameters\vspace*{1pt} in the Fama--French model, we will use the
two-year daily data $(\ty_t, \tf_t)$ from Jan $1$st, 2009 to Dec
$31$st, 2010 ($T=500$) of 30 industry portfolios.
\begin{longlist}[(3)]
\item[(1)] Calculate the least\vspace*{1pt} squares estimator $\tB$ of
$\ty_t=\bB \tf _t+\bu_ t$, and take the rows of $\tB$, namely\vadjust{\goodbreak}
$\tilde{\mathbf b} _1=(b_{11},b_{12},b_{13}),\ldots,\tilde{\mathbf
b}_{30}=(b_{30,1},b_{30,2},b_{30,3})$, to calculate the sample mean
vector $\bmu_B$ and sample covariance
%
\begin{table}
\tablewidth=185pt
\caption{Mean and covariance matrix used to generate $\mathbf{b}$}
\label{table1}
\begin{tabular*}{\tablewidth}{@{\extracolsep{\fill}}l@{\hspace*{30pt}}ccc@{}}
\hline
$\bolds{\bmu_B}$ & \multicolumn{3}{c@{}}{$\bolds{\Sig_B}$} \\
\hline
\hphantom{$-$}1.0641 &0.0475 & 0.0218 & 0.0488 \\
\hphantom{$-$}0.1233 &0.0218 & 0.0945 &0.0215 \\
$-$0.0119&0.0488 & 0.0215 & 0.1261 \\
\hline
\end{tabular*}
\end{table}
matrix $\Sig_B$. The results are depicted in Table \ref{table1}. We then create a
mutlivariate normal distribution $N_3(\bmu_B,\Sig_B)$, from which the
factor loadings $\{\mathbf{b}_i\}_{i=1}^p$ are drawn.
\item[(2)] For each fixed $p$, create the sparse matrix
$\Sig_u=\bD+\mathbf{s}{\mathbf s}'-\operatorname{diag}\{
s_1^2,\ldots,\allowbreak s_p^2\}$ in the following
way. Let $\hu_t=\ty_t-\tB\tf_t$. For $i=1,\ldots,30$, let $\widehat
\sigma_i$
denote the standard deviation of the residuals of the $i$th portfolio.
We find $\min(\hat\sigma_i)=0.3533$, $\max(\hat\sigma_i)=1.5222$ and
calculate the mean and the standard deviation of the $\widehat\sigma
_i$'s, namely
$\bar{\sigma}=0.6055$ and $\sigma_{\mathrm{SD}}=0.2621$.

Let $\bD=\operatorname{diag}\{\sigma_1^2,\ldots,\sigma_p^2\}$, where
$\sigma_1,\ldots,\sigma_p$ are generated independently from the Gamma
distribution $G(\alpha, \beta)$, with mean $\alpha\beta$ and standard
deviation~$\alpha^{1/2}\beta$. We match these values to $\bar\sigma
=0.6055$ and $\sigma_{\mathrm{SD}}=0.2621$, to get $\alpha=5.6840$ and $\beta
=0.1503$. Further, we create a loop that only accepts the value of
$\sigma_i$ if it is between $\min(\hat\sigma_i)=0.3533$ and $\max
(\hat
\sigma_i)=1.5222$.

Create\vspace*{-1pt} ${\mathbf s}=(s_1,\ldots, s_p)'$ to be a sparse vector. We set
each $s_i
\sim N(0,1)$ with probability $\frac{0.2}{\sqrt{p}\log{p}}$, and
$s_i=0$ otherwise. This leads to an average of $\frac{0.2\sqrt
{p}}{\log
{p}}$ nonzero elements per each row of the error covariance matrix.

Create a loop that generates $\Sig_u$ multiple times until it is
positive definite.

\item[(3)] Assume the factors follow the vector autoregressive
[VAR(1)] model $\bff_t=\bmu+\bPhi{\bff_{t-1}}+\beps_t$ for some
%
\begin{table}[b]
\caption{Parameters of $\bff_t$ generating process}\label{table2}
\begin{tabular*}{\tablewidth}{@{\extracolsep{\fill}}l@{\hspace*{30pt}}ccc@{\hspace*{30pt}}d{2.4}d{2.4}d{2.4}@{}}
\hline
$\bmu$ & \multicolumn{3}{c}{$\bolds{\cov(\bff_t)}$\hspace*{30pt}} & \multicolumn{3}{c@{}}{$\bPhi$}\\
\hline
0.1074 &2.2540 &0.2735&0.9197&-0.1149 &0.0024& 0.0776\\
0.0357 &0.2735 & 0.3767 &0.0430 &0.0016 & -0.0162 &0.0387 \\
0.0033&0.9197 &0.0430 &0.6822 &-0.0399 &0.0218 & 0.0351 \\
\hline
\end{tabular*}
\end{table}
$3\times3$ matrix $\bPhi$, where $\beps_t$'s are i.i.d. $N_3(0,\Sig
_{\epsilon})$. We estimate $\bPhi, \bmu$ and $\Sig_{\epsilon}$ from the
data, and obtain $\cov(\bff_t)$. They are summarized in Table
\ref{table2}.
\end{longlist}

\subsection{Simulation}
For each fixed $p$, we generate $(\mathbf{b}_1,\ldots,\mathbf{b}_p)$
independently from
$N_3(\bmu_B,\Sig_B)$, and generate $\{\bff_t\}_{t=1}^T$ and $\{\bu
_t\}
_{t=1}^T$ independently. We keep $T=500$ fixed, and gradually increase
$p$ from $20$ to $600$ in multiples of $20$ to illustrate the rates of
convergence when the number of variables diverges with respect to the
sample size.

Repeat the following steps $N=200$ times for each fixed $p$:
\begin{longlist}[(5)]
\item[(1)] Generate $\{\mathbf{b}_i\}_{i=1}^p$ independently from
$N_3(\bmu
_B,\Sig_B)$, and set $\bB=(\mathbf{b}_1,\allowbreak\ldots,\mathbf{b}_p)'$.
\item[(2)] Generate $\{\bu_t\}_{t=1}^T$ independently from
$N_p(0,\Sig_u)$.\vspace*{1pt}
\item[(3)] Generate $\{\bff_t\}_{t=1}^T$ independently from the VAR(1)
model $\bff_t=\bmu+\bPhi{\bff_{t-1}}+\beps_t$.
\item[(4)] Calculate $\by_t={\bB}\bff_t+\bu_t$ for $t=1,\ldots,T$.
\item[(5)] Set $\omega_T\,{=}\,0.10K\sqrt{\log p/T}$ to obtain the
thresholding estimator (\ref{e25})~$\hSig{}^{\mathcal{T}}_u$ and the
sample covariance matrices $\hcov(\bff_t)$, \mbox{$\hSig_y\,{=}\,\frac
{1}{T\,{-}\,1}\sum_{t=1}^{T}(\by_t\,{-}\,\bar{\mathbf{y}})(\by_t\,{-}\,\bar{\mathbf{y}})^T$}.
\end{longlist}

We graph the convergence of $\hSig{}^{\mathcal{T}}$ and $\hSig_y$ to
$\Sig$, the covariance matrix of $\mathbf{y}$, under the entropy-loss norm
\mbox{$\|\cdot\|_{\Sigma}$} and the elementwise norm \mbox{$\|\cdot\|_{\Max}$}. We also
graph the convergence of the inverses $(\hSig{}^{\mathcal{T}})^{-1}$ and
$\hSig_y^{-1}$ to $\Sig^{-1}$ under the operator norm. Note that we
graph that only for $p$ from 20 to 300. Since $T=500$, for $p>500$ the
sample covariance matrix is singular. Also, for $p$ close to $500$,
$\hSig_y$ is nearly singular, which leads to abnormally large values of
the operator norm. Last, we record the standard deviations of these
norms.

\subsection{Results}

\begin{figure}

\includegraphics{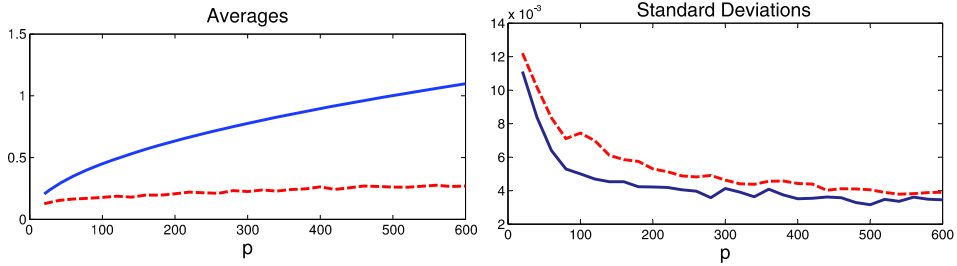}

\caption{Averages and standard deviations of
$\|\hSig{}^{\mathcal{T}}-\Sig\|_\Sigma$ (dashed curve) and
$\|\hSig_y-\Sig\|_\Sigma$ (solid curve) over $N=200$ iterations, as a
function of the dimensionality $p$.} \label{fig1}
\vspace*{-3pt}
\end{figure}

In Figures \ref{fig1}--\ref{fig3}, the dashed curves correspond to $\hSig{}^{\mathcal{T}}$
and the solid curves correspond to the sample covariance matrix $\hSig
_y$. Figures \ref{fig1} and \ref{fig2} present the averages and standard deviations of
%
the estimation error of both of these matrices with respect to the
$\Sigma$-norm and infinity norm, respectively. Figure \ref{fig3} presents the
averages and estimation errors of the inverses with respect to the
operator norm. Based on the simulation results, we can make the
following observations:
\begin{longlist}[(5)]
\item[(1)]
The standard deviations of the norms are negligible when compared to
their corresponding averages.
\item[(2)]
Under the $\|\cdot\|_{\Sigma}$, our estimate of the covariance matrix of
$\mathbf{y}$, $\hSig{}^{\mathcal{T}}$ performs much better than the sample
covaraince matrix $\hSig_y$. Note that,\vadjust{\goodbreak} in the proof of Theorem 2 in
\citet{FanFanLv08}, it was shown that
%
\begin{equation}\label{equ5.1}
\|\hSig_y- \Sig\|^{2}_{\Sigma}=O_p\biggl(\frac{K^3}{Tp}
\biggr)+O_p
\biggl(\frac{p}{T}\biggr)+O_p\biggl(\frac{K^{3/2}}{T}\biggr).
\end{equation}
For a small fixed value of $K$, such as $K=3$, the dominating term in
(\ref{equ5.1}) is~$O(\frac{p}{T})$. From Theorem \ref{t41}, and given that
%
\begin{figure}

\includegraphics{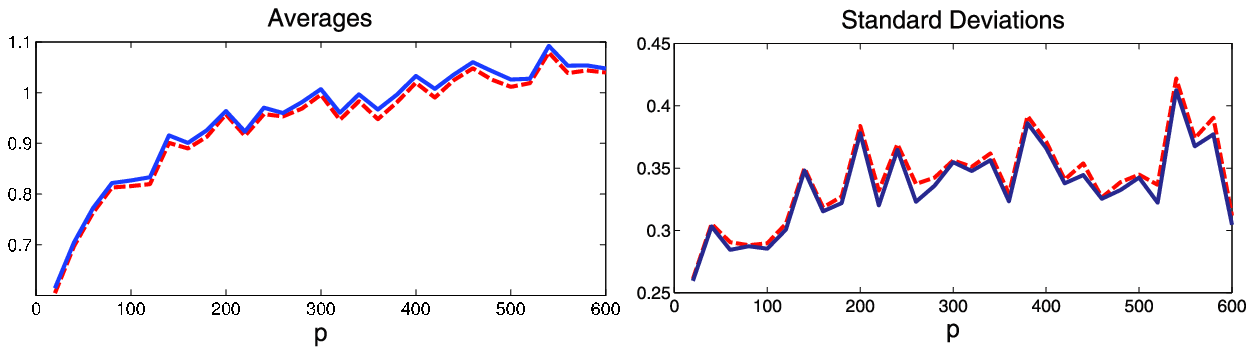}%
\vspace*{-2pt}
\caption{Averages and standard deviations of
$\|\hSig{}^{\mathcal{T}}-\Sig\|_{\Max}$ (dashed curve) and
$\|\hSig_y-\Sig\|_{\Max}$ (solid curve) over $N=200$ iterations, as a
function of the dimensionality $p$.}
\label{fig2}
\vspace*{-2pt}
\end{figure}
$m_T=o(p^{1/4})$, the dominating term in the convergence of $\|\hSig
{}^{\mathcal{T}}-\Sig\|^2_\Sigma$ is $O_p(\frac{p}{T^2}+\frac
{m_T^2\log p}{T})$. So, we would expect our estimator to perform
better, and the simulation results are consistent with the theory.
\item[(3)]
Under the infinity norm, both estimators perform roughly the same. This
is to be expected, given that the thresholding affects mainly the
%
\begin{figure}[b]
\vspace*{-2pt}
\includegraphics{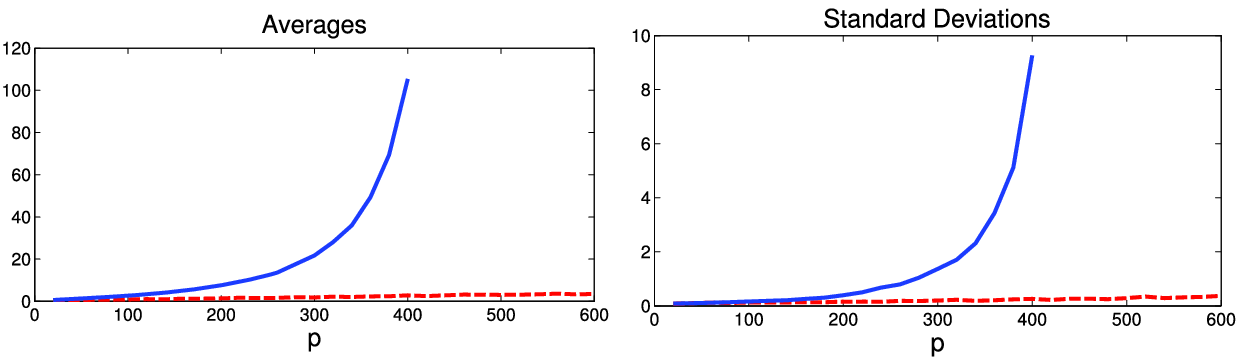}%
\vspace*{-2pt}
\caption{Averages and standard deviations of
$\|(\hSig{}^{\mathcal{T}})^{-1}-\Sig^{-1}\|$ (dashed curve) and
$\|\hSig_y^{-1}-\Sig^{-1}\|$ (solid curve) over $N=200$ iterations, as
a function of the dimensionality $p$.}
\label{fig3}
\end{figure}
elements of the covariance matrix that are closest to $0$, and the
infinity norm depicts the magnitude of the largest elementwise absolute error.
\item[(4)]
Under the operator norm, the inverse of our estimator, $(\hSig
{}^{\mathcal
{T}})^{-1}$ also performs significantly better than the inverse of the
sample covariance matrix.
\item[(5)] Finally, when $p>500$, the thresholding estimators $
\hSig{}^{\mathcal{T}}_u$ and $\hSig{}^{\mathcal{T}}$ are still
nonsingular.\vspace*{-2pt}\vadjust{\goodbreak}
\end{longlist}

In conclusion, even after imposing less restrictive assumptions on the
error covariance matrix, we still reach an estimator $\hSig{}^{\mathcal
{T}}$ that significantly outperforms the standard sample covariance
matrix.

\section{Conclusions and discussions}\label{sec6}

We studied the rate of convergence of high-dimensional covariance
matrix of approximate factor models under various norms. By assuming
sparse error covariance
matrix, we allow for the presence of the cross-sectional correlation
even after taking out common factors. Since direct observations of the
noises are not available, we constructed the error sample covariance
matrix, first based on the estimation residuals, and then estimated the
error covariance matrix using the adaptive thresholding method. We then
constructed the covariance matrix of $\by_t$ using the factor model,
assuming that the factors follow a stationary and ergodic process, but
can be weakly dependent. It was shown that after thresholding, the
estimated covariance matrices are still invertible even if $p>T$, and
the rate of convergence of $(\hSig{}^{\mathcal{T}})^{-1}$ and $(\hSig
{}^{\mathcal{T}}_u)^{-1}$ is of order $O_p(Km_T\sqrt{\log p/T})$, where~$K$
comes from the impact of estimating the unobservable noise terms.
This demonstrates when estimating the inverse covariance matrix, $p$ is
allowed to be much larger than $T$.

In fact, the rate of convergence in Theorem \ref{t21} reflects the
impact of unobservable idiosyncratic components on the thresholding
method. Generally, whether it is the minimax rate when direct
observations are not available but have to be estimated is an important
question, which is left as a research direction in the future.

Moreover, this paper uses the hard-thresholding technique, which takes
the form of $\widehat\sigma_{ij}(\sigma_{ij})=\sigma_{ij}I(|\sigma
_{ij}|>\theta
_{ij})$ for some pre-determined threshold $\theta_{ij}$. Recently,
\citet{RotLevZhu09} and \citet{CaiLiu11} studied a more general
thresholding function of \citet{AntFan01}, which admits the
form $\widehat\sigma_{ij}(\theta_{ij})=s(\sigma_{ij})$, and also
allows for
soft-thresholding. It is easy to apply the more general thresholding
here as well, and the rate of convergence of the resulting covariance
matrix estimators should be straightforward to derive.

Finally, we considered the case when common factors are observable, as
in \citet{FamFre92}. In some applications, the common factors are
unobservable and need to be estimated [\citet{Bai03}]. In that case, it is
still possible to consistently estimate the covariance matrices using
similar techniques as those in this paper. However, the impact of high
dimensionality on the rate of convergence comes also from the
estimation error of the unobservable factors. We plan to address this
problem in a separate paper.

\begin{appendix}\label{app}
\section{\texorpdfstring{Proofs for Section \lowercase{\protect\ref{sec2}}}{Proofs for Section 2}}\label{appA}

\subsection{Lemmas}
The following lemmas are useful to be proved first, in which we
consider the operator norm
$\|\bA\|^2=\lambda_{\max}(\bA'\bA)$.
%
\begin{lem} \label{la1} Let $\bA$ be an $m\times m$ random matrix,
$\bB
$ be an $m\times m$ deterministic matrix and both $\bA$ and $\bB$ are
semi-positive definite. If there exists a positive sequence $\{c_T\}
_{T=1}^{\infty}$ such that for all large enough $T$, $\lambda_{\min
}(\bB
)>c_T$. Then
\[
P\bigl(\lambda_{\min}(\bA)\geq0.5c_T\bigr)\geq P(\|\bA
-\bB\|\leq
0.5c_T)
\]
and
\[
P\biggl(\|\bA^{-1}-\bB^{-1}\|\leq\frac{2}{c_T^2}\|\bA-\bB\|
\biggr)\geq
P(\|\bA-\bB\|\leq0.5c_T).
\]
\end{lem}
\begin{pf}
For any $\bv\in\mathbb{R}^m$ such that $\|\bv\|=1$, under the
event $\|\bA-\bB\|\leq0.5c_T$,
\begin{eqnarray*}
\bv'\bA\bv&=&\bv'\bB\bv-\bv'(\bB-\bA)\bv\geq\lambda_{\min
}(\bB)-\|\bA
-\bB\|\\[-2pt]
&\geq&0.5c_T.
\end{eqnarray*}
Hence $\lambda_{\min}(\bA)\geq0.5c_T$.

In addition, still under the event $\|\bA-\bB\|\leq0.5c_T$,
\begin{eqnarray*}
\|\bA^{-1}-\bB^{-1}\|&=&\|\bA^{-1}(\bB-\bA)\bB^{-1}\|\\[-2pt]
&\leq& \lambda_{\min}(\bA)^{-1}\|\bA-\bB\|\lambda_{\min}(\bB
)^{-1}\\[-2pt]
&=&2c_T^{-1}\|\bA-\bB\|.
\end{eqnarray*}
\upqed\end{pf}
%
\begin{lem} \label{la12} Suppose that the random variables $Z_1, Z_2$
both satisfy the exponential-type tail condition: There exist $r_1$,
$r_2\in(0,1)$ and $b_1, b_2>0$, such that $\forall s>0$,
\[
P(|Z_i|>s)\leq\exp\bigl(1-(s/b_i)^{r_i}\bigr),\qquad  i=1,2.
\]
Then for some $r_3$ and $b_3>0$, and any $s>0$,
%
\begin{equation} \label{ea0}
P(|Z_1Z_2|>s)\leq\exp\bigl(1-(s/b_3)^{r_3}\bigr).
\end{equation}
\end{lem}
\begin{pf}
We have, for any $s>0$, $M=(sb_2^{r_2/r_1}/b_1)^{r_1/(r_1+r_2)}$,
$b=b_1b_2$ and $r=r_1r_2/(r_1+r_2)$,
\begin{eqnarray*}P(|Z_1Z_2|>s)&\leq& P(M|Z_1|>s)+P(|Z_2|>M)\\[-2pt]
&\leq&\exp\bigl(1-(s/b_1M)^{r_1}\bigr)+\exp\bigl(1-(M/b_2)^{r_2}\bigr)\\[-2pt]
&=&2\exp\bigl(1-(s/b)^r\bigr).
\end{eqnarray*}
Pick up an $r_3\in(0,r)$, and $b_3>\max\{(r_3/r)^{1/r}b, (1+\log
2)^{1/r}b\}$; then it can be shown that $F(s)=(s/b)^r-(s/b_3)^{r_3}$ is
increasing when $s>b_3$. Hence $F(s)>F(b_3)>\log2$ when $s>b_3$, which
implies when $s>b_3$,
\[
P(|Z_1Z_2|>s)\leq2\exp\bigl(1-(s/b)^r\bigr)\leq\exp\bigl(1-(s/b_3)^{r_3}\bigr).
\]
When $s\leq b_3$,
\[
P(|Z_1Z_2|>s)\leq1\leq\exp\bigl(1-(s/b_3)^{r_3}\bigr).
\]
\upqed\end{pf}
%
\begin{lem} \label{la2} Under the assumptions of Theorem \ref{t21},
there exists a constant $C_r>0$ that does not depend on $(p, T)$, such
that when $C>C_r$,

\begin{longlist}
\item
\[
\vspace*{-6pt}P\Biggl(\max_{i,j\leq p}\Biggl|\frac{1}{T}\sum_{t=1}^Tu_{it}u_{jt}-\sigma
_{ij}\Biggr|>C\sqrt{\frac{\log p}{T}}\Biggr)=O\biggl(\frac{1}{p^2}\biggr),
\]

\item
\[
\vspace*{-6pt}P\Biggl(\max_{i,j\leq p}\Biggl|\frac{1}{T}\sum_{t=1}^T(\hat{u}_{it}\hat
{u}_{jt}-u_{it}u_{jt})\Biggr|>Ca_T\Biggr)=O\biggl(\frac{1}{p^2}+\kappa
_1(p,T)\biggr),
\]

\item
\[
\vspace*{-6pt}P\Biggl(\max_{i,j\leq p}|\widehat\sigma_{ij}-\sigma_{ij}|>C\Biggl(\sqrt
{\frac{\log
p}{T}}+a_T\Biggr)\Biggr)=O\biggl(\frac{1}{p^2}+\kappa_1(p,T)\biggr).
\]
\end{longlist}
\end{lem}
\begin{pf}
(i) By Assumption \ref{a21} and Lemma \ref{la12}, $u_{it}u_{jt}$
satisfies the exponential tail condition, with parameter $r_1/3$ as
shown in the proof of Lemma \ref{la12}. Therefore by the Bernstein's
inequality [Theorem 1 of \citet{MerPelRio}], there exist constants
$C_1, C_2, C_3, C_4$ and $C_5>0$ that only depend on $b_1$, $r_1$ and
$r_2$ such that for any $i,j\leq p$, and $\gamma^{-1}=3r_1^{-1}+r_2^{-1}$,
\begin{eqnarray*}
P\Biggl(\Biggl|\frac{1}{T}\sum_{t=1}^Tu_{it}u_{jt}-\sigma_{ij}\Biggr|\geq s\Biggr)
&\leq& T\exp
\biggl(-\frac{(Ts)^{\gamma}}{C_1}\biggr)+\exp\biggl(-\frac
{T^2s^2}{C_2(1+TC_3)}\biggr)
\\
&&{}
+\exp\biggl(-\frac{(Ts)^2}{C_4T}\exp\biggl(\frac{(Ts)^{\gamma
(1-\gamma
)}}{C_5(\log Ts)^{\gamma}}\biggr)\biggr).
\end{eqnarray*}
Using Bonferroni's method, we have
\begin{eqnarray*}
&&
P\Biggl(\max_{i,j\leq p}\Biggl|\frac{1}{T}\sum_{t=1}^Tu_{it}u_{jt}-\sigma
_{ij}\Biggr|>s\Biggr)\\
&&\qquad\leq p^2\max_{i,j\leq p}P\Biggl(\Biggl|\frac{1}{T}\sum
_{t=1}^Tu_{it}u_{jt}-\sigma_{ij}\Biggr|>s\Biggr).
\end{eqnarray*}
Let $s=C\sqrt{(\log p)/T}$ for some $C>0$. It is not hard to check that
when $(\log p)^{2/\gamma-1}=o(T)$ (by assumption), for large enough $C$,
\[
p^2T\exp\biggl(-\frac{(Ts)^{\gamma}}{C_1}\biggr)+p^2\exp
\biggl(-\frac
{(Ts)^2}{C_4T}\exp\biggl(\frac{(Ts)^{\gamma(1-\gamma)}}{C_5(\log
Ts)^{\gamma}}\biggr)\biggr)=o\biggl(\frac{1}{p^2}\biggr)
\]
and
\[
p^2\exp\biggl(-\frac{T^2s^2}{C_2(1+TC_3)}\biggr)=O\biggl(\frac
{1}{p^2}\biggr).
\]
This proves (i).\vadjust{\goodbreak}

(ii) For some $C_1'>0$ such that
%
\begin{equation}\label{ea2}
P\Biggl(\max_{i\leq p}\frac{1}{T}\sum
_{t=1}^T(\hat
{u}_{it}-u_{it})^2> C_1'a_T^2\Biggr)=O(\kappa_1(p,T))\vspace*{-2pt}
\end{equation}
under the event $\{{\max_{i\leq p}}|\frac{1}{T}\sum
_{t=1}^Tu_{it}^2-\sigma
_{ii}|\leq\max_{i\leq p}\sigma_{ii}/4\}\cap\{\max_{i\leq p}\frac
{1}{T}\times\allowbreak\sum_{t=1}^T(\hat{u}_{it}-u_{it})^2\leq C_1'a_T^2\}$, by the
Cauchy--Schwarz inequality,
\begin{eqnarray*}
Z&\equiv&\max_{i,j\leq
p}\Biggl|\frac{1}{T}\sum_{t=1}^T(\hat{u}_{it}\hat{u}_{jt}-u_{it}u_{jt})\Biggr|\\[-3pt]
&\leq& \max_{i,j\leq p}\Biggl|\frac{1}{T}\sum_{t=1}^T(\hat
{u}_{it}-u_{it})(\hat{u}_{jt}-u_{jt})\Biggr|+2\max_{i,j\leq
p}\Biggl|\frac{1}{T}\sum_{t=1}^Tu_{it}(\hat{u}_{jt}-u_{jt})\Biggr|\\[-3pt]
&\leq& \max_{i\leq p}\frac{1}{T}\sum_{t=1}^T(\hat
{u}_{it}-u_{it})^2+2\sqrt{\max_{i\leq p}\frac{1}{T}\sum
_{t=1}^Tu_{it}^2}\sqrt{\max_{i\leq
p}\frac{1}{T}\sum_{t=1}^T(\hat{u}_{it}-u_{it})^2}\\[-3pt]
&\leq&C_1'a_T^2+2\sqrt{\frac{5}{4}\max_{i\leq p}\sigma_{ii}}\sqrt
{C_1'a_T^2}.\vspace*{-2pt}
\end{eqnarray*}
Since $a_T=o(1)$, when $C>3\sqrt{C_1'\max_{i\leq p}\sigma_{ii}}$, we
have, for all large $T$,
\[
Ca_T>C_1'a_T^2+2\sqrt{\frac{5}{4}\max_{i\leq p}\sigma_{ii}}\sqrt
{C_1'a_T^2}\vspace*{-2pt}
\]
and
\begin{eqnarray*}
P(Z\leq Ca_T)&\geq&1-P\Biggl(\max_{i\leq p}\Biggl|\frac{1}{T}\sum
_{t=1}^Tu_{it}^2-\sigma_{ii}\Biggr|>\max_{i\leq p}\sigma_{ii}/4\Biggr)\\[-3pt]
&&{}-P\Biggl(\max
_{i\leq p}\frac{1}{T}\sum_{t=1}^T(\hat{u}_{it}-u_{it})^2> C_1'a_T^2\Biggr).\vspace*{-2pt}
\end{eqnarray*}
By part (i) and (\ref{ea2}), $P(Z\leq Ca_T)\geq1-O(p^{-2}+\kappa
_1(p, T))$.

(iii)
By (i) and (ii), there exists $C_r>0$, when $C>C_r$, the displayed
inequalities in (i) and (ii) hold. Under the event $\{{\max_{i,j\leq
p}}|\frac{1}{T}\sum_{t=1}^Tu_{it}u_{jt}-\sigma_{ij}|\leq C\sqrt
{(\log
p)/T}\}\cap\{{\max_{i,j\leq p}}|\frac{1}{T}\sum_{t=1}^T\hat
{u}_{it}\hat
{u}_{jt}-u_{it}u_{jt}|\leq Ca_T\}$,
by the\vadjust{\goodbreak} triangular inequality,
\begin{eqnarray*}{\max_{i,j\leq p}}|\widehat\sigma_{ij}-\sigma
_{ij}|&\leq&\max
_{i,j\leq p}\Biggl|\frac{1}{T}\sum_{t=1}^Tu_{it}u_{jt}-\sigma_{ij}\Biggr|
+\max_{i,j\leq p}\Biggl|\frac{1}{T}\sum_{t=1}^T\hat{u}_{it}\hat{u}_{jt}-u_{it}u_{jt}\Biggr|\\[-3pt]
&\leq& C\Biggl(\sqrt{\frac{\log p}{T}}+a_T\Biggr).\vspace*{-2pt}
\end{eqnarray*}
Hence the desired result follows from part (i) and part (ii) of
the\vadjust{\goodbreak}
lemma.~%
\end{pf}
%
\begin{lem}\label{la3} Under Assumptions \ref{a21}, \ref{a22},
\[
P\Bigl(C_L\leq\min_{ij}\hat{\theta}_{ij}\leq\max_{ij}\hat{\theta
}_{ij}\leq
C_U\Bigr)\geq1-O\biggl(\frac{1}{p^2}+\kappa_1(p, T)+\kappa_2(p, T)\biggr),
\]
where
\begin{eqnarray*}
C_L&=&\frac{1}{45}\min_{ij}\var(u_{it}u_{jt}),\\
C_U&=&3\max_{i\leq p}\sigma_{ii}+4\max_{ij}\var(u_{it}u_{jt}).
\end{eqnarray*}
\end{lem}
\begin{pf}
(i) Using Bernstein's inequality and the same argument as in the proof
of Lemma \ref{la2}(i), there exists $C_r'>0$, when $C>C_r'$ and
$(\log
p)^{6/\gamma-1}=o(T)$,
\[
P\Biggl(\max_{i,j\leq p}\Biggl|\frac{1}{T}\sum_{t=1}^T(u_{it}u_{jt}-\sigma
_{ij})^2-\var(u_{it}u_{jt})\Biggr|>C\sqrt{\frac{\log p}{T}}\Biggr)=O
\biggl(\frac{1}{p^2}\biggr).\vadjust{\goodbreak}
\]
For some $C>0$, under the event $\bigcap_{i=1}^4A_i$, where
\begin{eqnarray*}
A_1&=&\Biggl\{{\max_{i,j\leq p}}|\sigma_{ij}-\widehat\sigma
_{ij}|\leq
C\Biggl(\sqrt{\frac{\log
p}{T}}+a_T\Biggr)\Biggr\},\\
A_2&=&\biggl\{{\max_{i\leq p, t\leq T}}|\hat{u}_{it}-u_{it}|\leq
\min\biggl\{\frac{1}{2},\sqrt{\Bigl(20\max_i\sigma_{ii}\Bigr)^{-1}\min_{ij}\var
(u_{it}u_{jt})}\biggr\}\biggr\},\\
A_3&=&\Biggl\{\max_{i\leq p}\Biggl|\frac{1}{T}\sum_{t=1}^Tu_{it}^2-\sigma
_{ii}\Biggr|\leq
C\sqrt{\frac{\log
p}{T}}\Biggr\},\\
A_4&=&\Biggl\{\max_{i,j\leq p}\Biggl|\frac{1}{T}\sum_{t=1}^T(u_{it}u_{jt}-\sigma
_{ij})^2-\var(u_{it}u_{jt})\Biggr|\leq C\sqrt{\frac{\log p}{T}}\Biggr\},
\end{eqnarray*}
we have, for any $i,j$, by adding and subtracting terms,
\begin{eqnarray*}
\hat{\theta}_{i,j}&=&\frac{1}{T}\sum_t(\hat
{u}_{it}\hat
{u}_{jt}-\widehat\sigma_{ij})^2\\
&\leq&
\frac{2}{T}\sum_t(\hat{u}_{it}\hat{u}_{jt}-\sigma_{ij})^2+2\max
_{i,j}(\sigma_{ij}-\widehat\sigma_{ij})^2\\
&\leq&
\frac{4}{T}\sum_t(\hat{u}_{it}-u_{it})^2\hat{u}_{jt}^2+\frac
{4}{T}\sum
_t(\hat{u}_{jt}-u_{jt})^2{u}_{it}^2\\
&&{}+\frac{4}{T}\sum
_t(u_{it}u_{jt}-\sigma_{ij})^2+O\biggl(\frac{\log p}{T}+a_T^2\biggr)\\
&\leq&{4\max_{it}}|\hat{u}_{it}-u_{it}|^2\biggl(\max_i\widehat\sigma
_{ii}+\max_i\frac
{1}{T}\sum_tu_{it}^2\biggr)\\
&&{}+4\var(u_{it}u_{jt})+O\Biggl(\sqrt{\frac{\log p}{T}}+\frac{\log p}{T}+a_T^2\Biggr)\\
&\leq&\Biggl(2C\sqrt{\frac{\log
p}{T}}+Ca_T+2\max_i\sigma_{ii}\Biggr)\\
&&{}+4\var(u_{it}u_{jt})+o(1),
\end{eqnarray*}
where the $O(\cdot)$ and $o(\cdot)$ terms are uniform in $p$ an $T$. Hence
under $\bigcap_{i=1}^4A_i$, for all large enough $T,p$, uniformly in
$i,j$, we have
\[
\hat{\theta}_{i,j}\leq3\max_{i\leq p}\sigma_{ii}+4\max_{ij}\var
(u_{it}u_{jt}).
\]

Still by adding and subtracting terms, we obtain
\begin{eqnarray*}
&&\frac{1}{T}\sum_t(u_{it}u_{jt}-\sigma_{ij})^2\\
&&\qquad\leq\frac{4}{T}\sum_t(u_{it}u_{jt}-\hat{u}_{it}\hat
{u}_{jt})^2+\frac
{4}{T}\sum_t(\hat{u}_{it}\hat{u}_{jt}-\widehat\sigma_{ij})^2
+4(\sigma_{ij}-\widehat\sigma_{ij})^2\\
&&\qquad\leq\frac{8}{T}\sum_tu_{it}^2(u_{jt}-\hat{u}_{jt})^2+\frac
{8}{T}\sum
_t\hat{u}_{jt}^2(u_{it}-\hat{u}_{it})^2+4\hat{\theta}_{ij}+O\biggl(\frac
{\log
p}{T}+a_T^2\biggr)\\
&&\qquad\leq{8\max_{it}}|\hat{u}_{it}-u_{it}|^2\biggl(\max_i\widehat\sigma
_{ii}+\max_{j}\frac
{1}{T}\sum_tu_{jt}^2\biggr)+4\hat{\theta}_{ij}+o(1).
\end{eqnarray*}
Under the event $\bigcap_{i=1}^4A_i$, we have
\begin{eqnarray*}
4\hat{\theta}_{ij}+o(1)&\geq& \min_{ij}\var
(u_{it}u_{jt})-C\sqrt{\frac{\log
p}{T}}\\
&&{}-8\max_{it}|\hat{u}_{it}-u_{it}|^2\Biggl[2C\sqrt{\frac{\log
p}{T}}+Ca_T+2\max_{i}\sigma_{ii}\Biggr]\\
&\geq&\frac{1}{10}\min_{ij}\var(u_{it}u_{jt}).
\end{eqnarray*}
Hence for all large $T,p$, uniformly in $i,j$, we have $\hat{\theta
}_{ij}\geq\frac{1}{45}\min_{ij}\var(u_{it}u_{jt})$.

Finally, by Lemma \ref{la2} and Assumption \ref{a22},
\[
P\Biggl(\bigcap_{i=1}^4A_i\Biggr)\geq1-O\biggl(\frac{1}{p^2}+\kappa_1(p,T)+\kappa_2(p,T)\biggr),
\]
which completes the proof.
\end{pf}

\subsection{\texorpdfstring{Proof of Theorem \protect\ref{t21}}{Proof of Theorem 2.1}}

(i) For the operator norm, we have
\[
\|\hSig{}^{\mathcal{T}}_u-\Sig_u\|\leq\max_{i\leq p}\sum
_{j=1}^p\bigl|\widehat\sigma
_{ij}I(|\widehat\sigma_{ij}|\geq\omega_T\hat{\theta
}_{ij}^{1/2})-\sigma_{ij}\bigr|.
\]
By Lemma \ref{la2}(iii), there exists $C_1>0$ such that the event
\[
A_1'=\Biggl\{\max_{i,j\leq p}|\widehat\sigma_{ij}-\sigma_{ij}|\leq
C_1\Biggl(\sqrt{\frac
{\log p}{T}}+a_T\Biggr)\Biggr\}
\]
occurs with probability $P(A_1')\geq1-O(\frac{1}{p^2}+\kappa
_1(p,T))$. Let $C>0$ be such that $C\sqrt{C_L}>2C_1$, where $C_L$
is defined in Lemma \ref{la3}. Let $\omega_T=C(\sqrt{\frac{\log
p}{T}}+a_T)$, $b_T=C_1(\sqrt{\frac{\log p}{T}}+a_T)$, then $\sqrt
{C_L}\omega_T>2b_T$, and by Lemma \ref{la3},
\begin{eqnarray*}
P\Bigl(\min_{ij}\hat{\theta}_{ij}^{1/2}\omega_T>2b_T\Bigr)&\geq&
P\Bigl(\min_{ij}\hat{\theta}_{ij}^{1/2}>\sqrt{C_L}\Bigr)\\
&\geq& 1-O\biggl(\frac{1}{p}+\kappa_1(p, T)+\kappa_2(p, T)\biggr).
\end{eqnarray*}
Define the following events:
\begin{eqnarray*}
A_2'&=&\Bigl\{\min_{ij}\hat{\theta}_{ij}^{1/2}\omega_T>2b_T\Bigr\},\\
A_3'&=&\Bigl\{\max_{ij}\hat{\theta}_{ij}^{1/2}\leq C_U^{1/2}\Bigr\},
\end{eqnarray*}
where $C_U$ is defined in Lemma \ref{la3}. Under $\bigcap_{i=1}^3A_i'$,
the event $|\widehat\sigma_{ij}|\geq\omega_T\hat{\theta
}_{ij}^{1/2}$ implies
$|\sigma_{ij}|\geq b_T$, and the event $|\widehat\sigma_{ij}|<\omega
_T\hat{\theta
}_{ij}^{1/2}$ implies $|\sigma_{ij}|<b_T+\sqrt{C_U}\omega_T$. We thus
have, uniformly in $i\leq p$, under $\bigcap_{i=1}^3A_i'$,
\begin{eqnarray*}
\|\hSig{}^{\mathcal{T}}_u-\Sig_u\|&\leq&\sum_{j=1}^p\bigl|\widehat\sigma
_{ij}I(|\widehat\sigma
_{ij}|\geq
\omega_T\hat{\theta}_{ij}^{1/2})-\sigma_{ij}\bigr|\\
&\leq& \sum_{j=1}^p|\widehat\sigma_{ij}-\sigma_{ij}|I(|\widehat
\sigma_{ij}|\geq
\omega_T\hat{\theta}_{ij}^{1/2})+\sum_{j=1}^p|\sigma
_{ij}|I(|\widehat\sigma
_{ij}|<\omega_T\hat{\theta}_{ij}^{1/2})\\
&\leq& \sum_{j=1}^p|\widehat\sigma_{ij}-\sigma_{ij}|I(|\sigma
_{ij}|\geq
b_T)+\sum_{j=1}^p|\sigma_{ij}|I\bigl(|\sigma_{ij}|<b_T+\sqrt{C_U}\omega
_T\bigr)\\
&\leq& b_Tm_T+\bigl(b_T+\sqrt{C_U}\omega_T\bigr)m_T\\
&\leq&\bigl(\sqrt{C_L}+\sqrt{C_U}\bigr)\omega_Tm_T.
\end{eqnarray*}
By Lemmas \ref{la2}(iii) and \ref{la3}, $P(\bigcap_{i=1}^3A_i')\geq
1-O(\frac{1}{p^2}+\kappa_1(p, T)+\kappa_2(p,T))$, which proves the result.

(ii)
By part (i) of the theorem, there exists some $C>0$,
\[
P(\|\hSig{}^{\mathcal{T}}_u-\Sig_u\|>C\omega_Tm_T)=O\biggl(\frac
{1}{p^2}+\kappa_1(p, T)+\kappa_2(p,T)\biggr).
\]
By Lemma \ref{la1},
\begin{eqnarray*}
P\bigl(\lambda_{\min}(\hSig{}^{\mathcal{T}}_u)\geq0.5\lambda_{\min
}(\Sig
_u)\bigr)&\geq& P\bigl(\|\hSig{}^{\mathcal{T}}_u-\Sig_u\|\leq
0.5\lambda_{\min}(\Sig_u)\bigr)\\
&\geq&1-O\biggl(\frac{1}{p^2}+\kappa_1(p, T)+\kappa_2(p,T)\biggr).
\end{eqnarray*}

In addition, when $\omega_Tm_T=o(1)$,
\begin{eqnarray*}
&&P\bigl(\|(\hSig{}^{\mathcal{T}}_u)^{-1}-\Sig_u^{-1}\|\leq
2\|\Sig_u^{-1}\|C\omega_Tm_T\bigr)\\
&&\qquad\geq P\bigl(\|(\hSig{}^{\mathcal{T}}_u)^{-1}-\Sig_u^{-1}\|\leq2\|
\Sig
_u^{-1}\|\cdot\|\hSig{}^{\mathcal{T}}_u-\Sig_u\|,
\|\hSig{}^{\mathcal{T}}_u-\Sig_u\|\leq
C\omega_Tm_T\bigr)\\
&&\qquad\geq P\bigl(\|(\hSig{}^{\mathcal{T}}_u)^{-1}-\Sig_u^{-1}\|\leq2\|
\Sig
_u^{-1}\|\cdot\|\hSig{}^{\mathcal{T}}_u-\Sig_u\|\bigr)\\
&&\qquad\quad{}-P( \|
\hSig{}^{\mathcal{T}}_u-\Sig_u\|>
C\omega_Tm_T)\\
&&\qquad\geq P\bigl( \|\hSig{}^{\mathcal{T}}_u-\Sig_u\|\leq0.5\lambda
_{\min
}(\Sig_u)\bigr)\\
&&\qquad\quad{}-O\biggl(\frac{1}{p^2}+\kappa_1(p,
T)+\kappa_2(p,T)\biggr)\\
&&\qquad\geq1-O\biggl(\frac{1}{p^2}+\kappa_1(p, T)+\kappa_2(p,T)\biggr),
\end{eqnarray*}
where the third inequality follows from Lemma \ref{la1} as well.

\section{\texorpdfstring{Proofs for Section \lowercase{\protect\ref{sec3}}}{Proofs for Section 3}}\label{appB}

\subsection{\texorpdfstring{Proof of Theorem \protect\ref{t31}}{Proof of Theorem 3.1}}

\begin{lem}\label{lb1} There exists $C_1>0$ such that:

\begin{longlist}
\item
\[
P\Biggl(\max_{i,j\leq K}\Biggl|\frac{1}{T}\sum
_{t=1}^Tf_{it}f_{jt}-Ef_{it}f_{jt}\Biggr|>C_1\sqrt{\frac{\log T}{T}}
\Biggr)=O\biggl(\frac{1}{T^2}\biggr),
\]
\item
\[
P\Biggl(\max_{k\leq K, i\leq p}\Biggl|\frac{1}{T}\sum
_{t=1}^Tf_{kt}u_{it}\Biggr|>C_1\sqrt{\frac{\log p}{T}}\Biggr)=O
\biggl(\frac
{1}{p^2}\biggr).
\]
\end{longlist}
\end{lem}
\begin{pf}
(i) Let $Z_{ij}=\frac{1}{T}\sum_{t=1}^T(f_{it}f_{jt}-Ef_{it}f_{jt})$.
We bound $\max_{ij}|Z_{ij}|$ using a Bernstein-type inequality. Lemma
\ref{la12} implies that for any $i$ and \mbox{$j\leq K$}, $f_{it}f_{jt}$
satisfies the exponential tail condition (\ref{e333}) with parameter
$r_3/3$. Let $r_4^{-1}=3r_3^{-1}+r_2^{-1}$, where $r_2>0$ is the
parameter in the strong mixing condition. By Assumption \ref{a33},
$r_4<1$, and by the Bernstein inequality for weakly dependent data in
Merlev{\`e}de, Peligrad and Rio [(\citeyear{MerPelRio}), Theorem 1], there exist $C_i>0$, $i=1,\ldots,5$, for
any $s>0$
%
\begin{eqnarray} \label{eb1}
\max_{i,j}P(|Z_{ij}|>s)&\leq&
T\exp\biggl(-\frac{(Ts)^{r_4}}{C_1}\biggr)+\exp\biggl(-\frac
{T^2s^2}{C_2(1+TC_3)}\biggr)\nonumber\\[-8pt]\\[-8pt]
&&{}+\exp\biggl(-\frac{(Ts)^2}{C_4T}\exp\biggl(\frac
{(Ts)^{r_4(1-r_4)}}{C_5(\log Ts)^{r_4}}\biggr)\biggr).\nonumber
\end{eqnarray}
Using the Bonferroni inequality,
\[
P\Bigl(\max_{i\leq K, j\leq K}|Z_{ij}|>s\Bigr)\leq K^2 \max_{i,j}P(|Z_{ij}|>s).
\]
Let $s=C\sqrt{(\log T)/T}$. For all large enough $C$, since $K^2=o(T)$,
\begin{eqnarray*}
TK^2\exp\biggl(-\frac{(Ts)^{r_4}}{C_1}\biggr)+K^2\exp\biggl(-\frac
{(Ts)^2}{C_4T}\exp\biggl(\frac{(Ts)^{r_4(1-r_4)}}{C_5(\log
Ts)^{r_4}}\biggr)\biggr)&=&o\biggl(\frac{1}{T^2}\biggr),
\\
K^2\exp\biggl(-\frac{T^2s^2}{C_2(1+TC_3)}\biggr)&=&O\biggl(\frac
{1}{T^2}\biggr).
\end{eqnarray*}
This proves part (i).

(ii) By Lemma \ref{la12}, and Assumptions \ref{a21}(iii) and \ref
{a33}(ii), $Z'_{ki,t}\equiv f_{kt}u_{it}$ satisfies the exponential
tail condition (\ref{e251}) for the tail parameter
$2r_1r_3/(3r_1+3r_3)$, as well as the strong mixing condition with
parameter $r_2$. Hence again we can apply the Bernstein inequality for
weakly dependent data in Merlev{\`e}de, Peligrad and Rio
[(\citeyear{MerPelRio}), Theorem 1] and the Bonferroni method on
$Z'_{ki,t}$ similar to (\ref{eb1}) with the parameter
$\gamma_2^{-1}=1.5r_1^{-1}+1.5r_3^{-1}+r_2^{-1}$. It follows from
$3r_1^{-1}+r_2^{-1}>1$ and $3r_3^{-1}+r_2^{-1}>1$ that $\gamma _2<1$.
Thus when $s=C\sqrt{(\log p)/T}$ for large enough $C$, the term
\[
pK\exp\biggl(-\frac{T^2s^2}{C_2(1+TC_3)}\biggr)\leq p^{-2},
\]
and the rest terms on the right-hand side of the inequality, multiplied
by~$pK$ are of order $o(p^{-2})$. Hence when $(\log p)^{2/\gamma
_2-1}=o(T)$ (which is implied by the theorem's assumption) and
$K=o(p)$, there exists $C'>0$,
%
\begin{equation}\label{eab1}
P\Biggl(\max_{k\leq K, i\leq p}\Biggl|\frac{1}{T}\sum
_{t=1}^Tf_{kt}u_{it}\Biggr|>C'\sqrt{\frac{\log p}{T}}\Biggr)=O\biggl(\frac
{1}{p^2}\biggr).
\end{equation}
\upqed\end{pf}
\begin{pf*}{Proof of Lemma \ref{l31}}
(i) Since $K\sqrt{\log T}=o(\sqrt{T})$, and $\lambda_{\min}(E\bff
_t\bff
_t')$ is bounded away from zero, for large enough $T$, by Lemma \ref{lb1}(i),
%
\begin{eqnarray}\label{ebb3}
&&P\biggl(\biggl\|\frac{1}{T}\bX\bX'-E\bff_t\bff_t'\biggr\|\leq
0.5\lambda_{\min}(E\bff_t\bff_t')\biggr)\nonumber\\
&&\qquad\geq P\Biggl(K\max_{i\leq K, j\leq K}\Biggl|\frac{1}{T}\sum
_{t=1}^Tf_{it}f_{jt}-Ef_{it}f_{jt}\Biggr|\leq0.5\lambda_{\min}(E\bff
_t\bff
_t')\Biggr)\\
&&\qquad\geq 1-O\biggl(\frac{1}{T^2}\biggr).\nonumber
\end{eqnarray}
Hence by Lemma \ref{la1},
%
\begin{equation}\label{ebb4}
P\bigl(\lambda_{\min}(T^{-1}\bX\bX')\geq0.5\lambda_{\min}(E\bff
_t\bff
_t')\bigr)\geq1-O\biggl(\frac{1}{T^2}\biggr).
\end{equation}
As $\widehat{\mathbf b}_i-{\mathbf b}_i=(\bX\bX')^{-1}\bX
\bu_i$, we have $\|\widehat{\mathbf b}_i-\mathbf{b}_i\|
^2=\bu_i'\bX'(\bX\bX')^{-2}\bX\bu_i$. For \mbox{$C'>0$} such that (\ref{eab1})
holds, under the event
\[
A\,{\equiv}\,\Biggl\{\max_{k\leq K, i\leq p}\Biggl|\frac{1}{T}\sum
_{t=1}^Tf_{kt}u_{it}\Biggr|\,{\leq}\,C'\sqrt{\frac{\log p}{T}}\Biggr\}\,{\cap}\,\{\lambda
_{\min}(T^{-1}\bX\bX')\,{\geq}\,0.5\lambda_{\min}(E\bff_t\bff_t')\},
\]
we have
\begin{eqnarray*}
\|\widehat{\mathbf b}_i-\mathbf{b}_i\|^2&\leq&
\frac{4}{\lambda_{\min}(E\bff_t\bff_t')^2}\sum_{k=1}^K\Biggl(\frac
{1}{T}\sum
_{t=1}^Tf_{kt}u_{it}\Biggr)^2\\
&\leq&\frac{4K}{\lambda_{\min}(E\bff_t\bff_t')^2}\max_{k\leq K,
i\leq
p}\Biggl(\frac{1}{T}\sum_{t=1}^Tf_{kt}u_{it}\Biggr)^2\\
&\leq&\frac{4KC'^2\log p}{\lambda_{\min}(E\bff_t\bff_t')^2T}.
\end{eqnarray*}
The desired result then follows from that $P(A)\geq1-O(\frac
{1}{T^2}+\frac{1}{p^2})$.

(ii) For $C>\max_{i\leq K}Ef_{it}^2$, we have, by Lemma \ref{lb1}(i),
\begin{eqnarray*}
P\biggl(\frac{1}{T}\sum_t\|\bff_t\|^2>CK\biggr)&\leq& P
\Biggl(K\max
_{k\leq K}\Biggl|\frac{1}{T}\sum_{t=1}^Tf_{kt}^2-Ef_{kt}^2\Biggr|+K\max_{k\leq
K}Ef_{kt}^2>CK\Biggr)\\
&=&O\biggl(\frac{1}{T^2}\biggr).
\end{eqnarray*}
The result then follows from
\[
\max_{i\leq p}\frac{1}{T}\sum_{t=1}^T|u_{it}-\hat{u}_{it}|^2\leq
\max
_{i\leq p}\frac{1}{T}\sum_t\|\bff_t\|^2\|\widehat{\mathbf b}_i-\mathbf{b}_i\|^2
\]
and part (i).\vadjust{\goodbreak}

(iii) By Assumption \ref{a33}, for any $s>0$,
\begin{eqnarray*}
P\Bigl({\max_{t\leq T}}\|\bff_t\|>s\Bigr)&\leq& TP(\|\bff_t\|>s)\leq TK\max
_{k\leq
K}P(f_{kt}^2>s^2/K)\\
&\leq& TK\exp\biggl(-\biggl(\frac{s}{b_2\sqrt{K}}
\biggr)^{r_3}\biggr).
\end{eqnarray*}
When $s\geq C\sqrt{K}(\log T)^{1/r_3}$ for large enough $C$, that is,
$C^{r_3}>4b_2^{r_3}$,
\[
P\Bigl(\max_{t\leq T}\|\bff_t\|>C\sqrt{K}(\log T)^{1/r_3}\Bigr)\leq T^{-2}.
\]

The result then follows from
\[
{\max_{t\leq T, i\leq p}}|u_{it}-\hat{u}_{it}|={\max_{t\leq T, i\leq
p}}|(\widehat{\mathbf b}_i-\mathbf{b}_i)'\bff_t|\leq{\max_i}\|
\widehat{\mathbf b}_i-\mathbf{b}_i{\|\max_t}\|\bff_t\|
\]
and Lemma \ref{l31}(i).
\end{pf*}
\begin{pf*}{Proof of Theorem \ref{t31}}
Theorem \ref{t31} follows immediately from Theorem~\ref{t21} and Lemma
\ref{l31}.
\end{pf*}

\subsection{\texorpdfstring{Proof of Theorem \protect\ref{t32}, part (i)}{Proof of Theorem 3.2, part (i)}}
Define
\begin{eqnarray*}
\bD_T&=&\hcov({\bff_t})-\cov({\bff_t}),\qquad {\bC_T}=\hB
-{\bB} ,
\\
\bE&=&(\bu_1,\ldots,\bu_T).
\end{eqnarray*}
We have
%
\begin{eqnarray}\label{eb3}
\|\hSig{}^{\mathcal{T}}-\Sig\|_{\Sigma}^{2}&\leq&
4\|{\bB}\bD_T{\bB}'\|^2_{\Sigma}+24\|{\bB}\hcov(\bff){\bC_T}'\|
^2_{\Sigma}\nonumber\\[-8pt]\\[-8pt]
&&{}+16\|{\bC_T}\hcov(\bff){\bC_T}'\|^2_{\Sigma}+2\|
\hSig{}^{\mathcal{T}}_u-\Sig_u\|_{\Sigma}^{2}.\nonumber
\end{eqnarray}
We bound the terms on the right-hand side in the following lemmas.
%
\begin{lem}\label{lb2} There exists $C>0$, such that:

\begin{longlist}
\item
\[
P\biggl(\|\bD_T\|_F^2>\frac{CK^2\log T}{T}\biggr)=O(T^{-2});
\]
\item
\[
P\biggl(\|\bC_T\|_F^2>\frac{CKp\log
p}{T}\biggr)=O(T^{-2}+p^{-2}).
\]
\end{longlist}
\end{lem}
\begin{pf}
(i) Similarly to the proof of Lemma \ref{lb1}(i), it can be shown that
there exists $C_1>0$,
\[
P\Biggl(\max_{i\leq K}\Biggl|\frac{1}{T}\sum
_{t=1}^Tf_{it}-Ef_{it}\Biggr|>C_1\sqrt
{\frac{\log T}{T}}\Biggr)=O(T^{-2}).\vadjust{\goodbreak}
\]
Hence $\sup_K\max_{i\leq K}E|f_{it}|<\infty$ implies that there exists
$C>0$ such that
\[
P\Biggl(\max_{i, j\leq K}\Biggl|\frac{1}{T}\sum_{t=1}^Tf_{it}\frac
{1}{T}\sum
_{t=1}^Tf_{jt}-Ef_{it}Ef_{jt}\Biggr|>C\sqrt{\frac{\log T}{T}}\Biggr)=O(T^{-2}).
\]
The result then follows from Lemma \ref{lb1}(i) and that
\begin{eqnarray*}
\|\bD_T\|_F^2&\leq& K^2\Biggl(\max_{i, j\leq K}\Biggl|\frac{1}{T}\sum
_{t=1}^Tf_{it}f_{jt}-Ef_{it}f_{jt}\Biggr|^2\\
&&\qquad\hspace*{0pt}{}+\max_{i, j\leq K}\Biggl|\frac
{1}{T}\sum
_{t=1}^Tf_{it}\frac{1}{T}\sum_{t=1}^Tf_{jt}-Ef_{it}Ef_{jt}\Biggr|^2\Biggr).
\end{eqnarray*}

(ii) We have $\bC_T=\bE\bX'(\bX\bX')^{-1}$. By Lemma \ref{lb1}(ii),
there exists $C'>0$ such that
\[
P\Biggl(\max_{k, i}\Biggl|\frac{1}{T}\sum_{t=1}^Tf_{kt}u_{it}\Biggr|>C'\sqrt
{\frac
{\log p}{T}}\Biggr)=O(p^{-2}).
\]
Under the event
\[
A=\Biggl\{\max_{k, i}\Biggl|\frac{1}{T}\sum_{t=1}^Tf_{kt}u_{it}\Biggr|\leq C'\sqrt
{\frac
{\log p}{T}}\Biggr\}\cap\{\lambda_{\min}(T^{-1}\bX\bX')\geq0.5\lambda
_{\min
}(E\bff_t\bff_t')\},
\]
$\|\bC_T\|_F^2\leq4\lambda_{\min}^{-2}(E\bff_t\bff_t')C'^2pK(\log
p)/T$, which proves the result since\break $\lambda_{\min}(E\bff_t\bff_t')$
is bounded away from zero and $P(A)\geq1-O(T^{-2}+p^{-2})$ due to
(\ref{ebb4}).
\end{pf}
%
\begin{lem}\label{lb3} There exists $C>0$ such that:
\begin{longlist}
\item
\begin{eqnarray*}
&&P\biggl(\|{\bB}\bD_T{\bB}'\|^2_{\Sigma}+\|{\bB} \hcov(\bff_t)\bC
_T'\|
^2_{\Sigma}>\frac{CK\log p}{T}+\frac{CK^2\log T}{Tp}
\biggr)\\
&&\qquad=O(T^{-2}+p^{-2});
\end{eqnarray*}
\item
\[
P\biggl(\|{\bC_T}\hcov(\bff){\bC_T}'\|^2_{\Sigma}>\frac
{CpK^2(\log
p)^2}{T^2}\biggr)=O(T^{-2}+p^{-2}).
\]
\end{longlist}
\end{lem}

\begin{pf}
(i) The same argument in Fan, Fan and Lv [(\citeyear{FanFanLv08}),
proof of Theorem~2] implies that
\[
\|{\bB}'\Sig^{-1}{\bB}\|\leq2\|\cov(\bff_t)^{-1}\|=O(1).
\]
Hence
%
\begin{eqnarray}\label{eb4}
\|{\bB}\bD_T{\bB}'\|^2_{\Sigma}&=&p^{-1}\tr(\Sig^{-1/2}\bB\bD
_T\bB'\Sig
^{-1}\bB\bD_T\bB'\Sig^{-1/2})\nonumber\\
&=&p^{-1}\tr(\bD_T\bB'\Sig^{-1}\bB\bD_T\bB'\Sig^{-1}\bB)\nonumber\\[-8pt]\\[-8pt]
&\leq& p^{-1}\|\bD_T{\bB}'\Sig^{-1}{\bB}\|_F^2\nonumber\\
&\leq&O(p^{-1})\|\bD_T\|_F^2.\nonumber
\end{eqnarray}

On the other hand,
%
\begin{equation}
\|{\bB}\hcov(\bff){\bC_T}'\|^2_{\Sigma}\leq
8T^{-2}\|\bB\bX\bX'{\bC_T}'\|^2_{\Sigma}+8T^{-4}\|\bB\bX\bone
\bone'\bX
'{\bC_T}'\|_{\Sigma}^2.
\end{equation}
Respectively,
%
\begin{eqnarray}
\|\bB\bX\bX'{\bC_T}'\|^2_{\Sigma}&\leq& p^{-1}\|\bX\bX'\bC
_T'\Sig
^{-1}\|_F\|\bC_T\bX\bX
'{\bB}'\Sig^{-1}{\bB}\|_F,\nonumber\\[-8pt]\\[-8pt]
\|\bB\bX\bone\bone'\bX'\bC_T'\|_{\Sigma}^2&\leq& p^{-1}\|\bX
\bone\bone
'\bX'\bC_T'\Sig^{-1}\|_F\|\bC_T\bX\bone\bone'\bX
'{\bB}'\Sig^{-1}{\bB}\|_F.\nonumber
\end{eqnarray}

By Lemma \ref{lb1}(i), and $E\bff_t\bff_t'<\infty$, $P(\|\bX\bX'\|
>TC)=O(T^{-2})$ for some \mbox{$C>0$}. Hence, Lemma \ref{lb2}(ii) implies
%
\begin{equation}
P(\|\bB\bX\bX'{\bC_T}'\|^2_{\Sigma}>C'TK\log p)=O(T^{-2}+p^{-2})
\end{equation}
for some $C'>0$. In addition, the eigenvalues of $\hcov(\bff
_t)=T^{-1}\bX\bX'-\break T^{-2}\bX\bone\bone'\bX'$ are all bounded away from
both zero and infinity with probability at least $1-O(T^{-2})$ [implied
by Lemmas \ref{lb1}(i), \ref{la1} and Assumption~\ref{a34}]. Hence for
some $C_1>0$, with probability ast least $1-O(T^{-2})$,
%
\begin{eqnarray}\label{eb8}
\|\bX\bone\bone'\bX'\|&\leq&\|T\bX\bX'\|\leq
T^2C_1,\nonumber\\[-8pt]\\[-8pt]
\|\bB\bX\bone\bone'\bX'\bC_T'\|_{\Sigma}^2&\leq& O(p^{-1})
\|\bX\bone\bone'\bX'\|^2\|\bC_T\|_F^2.\nonumber
\end{eqnarray}
The result then follows from the combination of (\ref{eb4})--(\ref
{eb8}) and Lemma~\ref{lb2}.

(ii) Straightforward calculation yields
\begin{eqnarray*}
p\|{\bC_T}\hcov(\bff){\bC_T}'\|^2_{\Sigma}&=&\tr({\bC_T}\hcov
(\bff){\bC
_T}'\Sig^{-1}{\bC_T}\hcov(\bff){\bC_T}'\Sig^{-1})\\
&\leq& \|\bC_T\hcov(\bff){\bC_T}'\Sig^{-1}\|_F^2\\
&\leq& \lambda_{\max}^2(\Sig^{-1})\lambda_{\max}^2(\hcov(\bff
_t))\|\bC
_T\|_F^4.
\end{eqnarray*}

Since $\|\cov(\bff_t)\|$ is bounded, by Lemma \ref{lb1}(i), $\lambda
_{\max}^2(\hcov(\bff_t))$ is bounded with probability at least
$1-O(T^{-2})$. The result again follows from Lem\-ma~\ref{lb2}(ii).
\end{pf}
\begin{pf*}{Proof of Theorem \ref{t32}, part \textup{(i)}}
(a) We have
%
\begin{eqnarray}\label{eb9}
\|\hSig{}^{\mathcal{T}}_u-\Sig_u\|_{\Sigma}&=&p^{-1/2}\|\Sig
^{-1/2}(\hSig{}^{\mathcal{T}}_u-\Sig_u)\Sig^{-1/2}\|_F\nonumber\\
&\leq& \|\Sig^{-1/2}(\hSig{}^{\mathcal{T}}_u-\Sig_u)\Sig^{-1/2}\|\\
&\leq&
\|\hSig{}^{\mathcal{T}}_u-\Sig_u\|\cdot\lambda_{\max}(\Sig
^{-1}).\nonumber
\end{eqnarray}
Therefore, (\ref{eb3}), (\ref{eb9}), Theorem \ref{t31} and Lemmas
\ref
{lb2}, \ref{lb3} yield the result, with the fact that [assuming $\log
T=o(p)$]
\begin{eqnarray*}
&&
\frac{K\log p}{T}+\frac{K^2\log T}{Tp}+\frac{pK^2(\log
p)^2}{T^2}+\frac
{m_T^2K^2\log p}{T}\\
&&\qquad=O\biggl(\frac{pK^2(\log p)^2}{T^2}+\frac
{m_T^2K^2\log p}{T}\biggr).
\end{eqnarray*}

(b) For the infinity norm, it is straightforward to find that
%
\begin{eqnarray}\label{eb10}\qquad
\|\hSig{}^{\mathcal{T}}-\Sig\|_{\Max}&\leq&
\|2\bC_T\cov(\bff_t)\bB'\|_{\Max}+\|\bB\bD_T\bB'\|_{\Max}\nonumber\\
&&{}+\|\bC
_T\cov(\bff_t)\bC_T'\|_{\Max}+\|2\bB\bD_T\bC_T'\|_{\Max}\\
&&{}+\|\bC_T\bD_T\bC_T'\|_{\Max}
+\|\hSig{}^{\mathcal{T}}_u-\Sig_u\|_{\Max}.\nonumber
\end{eqnarray}
By assumption, both $\|\bB\|_{\Max}$ and $\|{\cov}(\bff_t)\|_{\Max}$
are bounded uniformly in $(p, K, T)$. In addition, let $\mathbf{e}_i$
be a $p$-dimensional column vector whose $i$th component is one with
the remaining components being zeros. Then under the events
$\|\bD_{T}\|_{\Max} \leq C\sqrt{(\log T)/T}$, ${\max_{i\leq K, j\leq
p}}|\frac {1}{T}\sum_{t=1}^Tf_{it}u_{jt}|\leq C\sqrt{(\log p)/T}$,
$\|\frac{1}{T}\bX\bX'\|\leq C$, and ${\max_{i\leq
p}}\|\widehat{\mathbf b}_i- \mathbf{b}_i\|\leq C\sqrt {K(\log p)/T}$,
we have, for some $C'>0$,
%
\begin{eqnarray}
\|2\bC_T\cov(\bff_t)\bB'\|_{\Max}&\leq&2\max_{i,j\leq
p}\|\mathbf{e}_i'\bC_T\cov(\bff_t)\bB'\mathbf{e}_j\|\nonumber
\\
&\leq& 2\max_{i\leq p}\|\widehat{\mathbf b}_i-\mathbf{b}_i\|\|
{\cov}(\bff_t){\|\max_{j\leq
p}}\|\mathbf{b}_j\|\\
&\leq& C'K\sqrt{\frac{\log p}{T}},\nonumber\\
\|\bC_T\|_{\Max}&=&\max_{i,j\leq p}\biggl|\mathbf{e}_i'\frac
{1}{T}\bE\bX'\biggl(\frac
{1}{T}\bX\bX'\biggr)^{-1}\mathbf{e}_j\biggr|\nonumber\\
&\leq&\max_{i\leq
p}\biggl\|\mathbf{e}_i'\frac{1}{T}\bE\bX'\biggr\|\cdot
\biggl\|\biggl(\frac{1}{T}\bX\bX
'\biggr)^{-1}\biggr\|
\nonumber\\[-8pt]\\[-8pt]
&\leq&\sqrt{K}\max_{i\leq K, j\leq
p}\Biggl|\frac{1}{T}\sum_{t=1}^Tf_{it}u_{jt}\Biggr|\cdot
\biggl\|\biggl(\frac{1}{T}\bX\bX
'\biggr)^{-1}\biggr\|\nonumber\\
&\leq& C'\sqrt{K(\log p)/T},\nonumber\\
\|\bB\bD_T\bB'\|_{\Max}&\leq& K^2 \|\bB\|_{\Max}^2\|\bD_T\|_{\Max}
\leq C'K^2\sqrt{\frac{\log T}{T}},
\\
\|\bC_T\cov(\bff_t)\bC_T'\|_{\Max}&\leq&\max_{i,j\leq
p}\|\mathbf{e}_i'\bC_T\cov(\bff_t)\bC_T'\mathbf{e}_j\|\nonumber\\
&\leq&\max_{i\leq p}\|\mathbf{e}_i'\bC_T\|^2\|{\cov}(\bff_t)\|
\\
&\leq&\frac{C'K^2\log
p}{T},\nonumber
\\
\|2\bB\bD_T\bC_T'\|_{\Max}&\leq&
2K^2\|\bB\|_{\Max}\|\bD_T\|_{\Max}\|\bC_T\|_{\Max}\nonumber\\[-8pt]\\[-8pt]
&=&o\Biggl(K^2\sqrt{\frac{\log T}{T}}\Biggr)\nonumber
\end{eqnarray}
and
%
\begin{equation}
\|\bC_T\bD_T\bC_T'\|_{\Max}\leq K^2\|\bD_T\|_{\Max}\|\bC_T\|_{\Max}^2
=o\Biggl(K^2\sqrt{\frac{\log T}{T}}\Biggr).
\end{equation}

Moreover, the $(i,j)$th entry of $\hSig{}^{\mathcal{T}}_u-\Sig_u$ is
given by
\[
\widehat\sigma_{ij}I\bigl(|\widehat\sigma_{ij}|\geq\omega_T\sqrt{\hat
{\theta}_{ij}}\bigr)-\sigma
_{ij}=\cases{
-\sigma_{ij}, &\quad if $|\widehat\sigma_{ij}|<\omega_T\sqrt
{\hat{\theta
}_{ij}}$, \cr
\widehat\sigma_{ij}-\sigma_{ij}, &\quad o.w.}
\]
Hence $\|\hSig{}^{\mathcal{T}}_u-\Sig_u\|_{\Max}\leq\max_{i,
j\leq
p}|\sigma_{ij}-\widehat\sigma_{ij}|+\omega_T\max_{i,j\leq p}\sqrt
{\hat{\theta
}_{ij}}$, which implies that with probability at least $1-O(p^{-2}+T^{-2})$,
%
\begin{equation} \label{eb17}
\|\hSig{}^{\mathcal{T}}_u-\Sig_u\|_{\Max}\leq C'K\sqrt{\frac{\log p}{T}}.
\end{equation}
The result then follows from the combination of (\ref{eb10})--(\ref
{eb17}), (\ref{ebb4}) and Lemmas~\ref{l31}, \ref{lb1}.
\end{pf*}

\subsection{\texorpdfstring{Proof of Theorem \protect\ref{t32}, part (ii)}{Proof of Theorem 3.2, part (ii)}}

We first prove two technical lemmas to be used below.
%
\begin{lem} \label{lb4}
\textup{(i)} $\lambda_{\min}({\bB}'\Sig_u^{-1}{\bB})\geq cp$ for some
$c>0$.\vspace*{1pt}

\textup{(ii)} $\|[\cov(\bff)^{-1}+{\bB}'\Sig_u^{-1}{\bB}]^{-1}\|=O(p^{-1})$.
\end{lem}
\begin{pf}
(i) We have
\[
\lambda_{\min}({\bB}'\Sig_u^{-1}{\bB})\geq\lambda_{\min}(\Sig
_u^{-1})\lambda_{\min}(\bB'\bB).
\]
It then follows from Assumption \ref{a35} that $\lambda_{\min}(\bB
'\bB
)>cp$ for some $c>0$ and all large $p$. The result follows since $\|
\Sig
_u\|$ is bounded away from infinity.

(ii) It follows immediately from
\[
\lambda_{\min}\bigl(\cov(\bff_t)^{-1}+{\bB}'\Sig_u^{-1}{\bB}\bigr)\geq
\lambda
_{\min}({\bB}'\Sig_u^{-1}{\bB}).
\]
\upqed\end{pf}
%
\begin{lem} \label{lb5} There exists $C>0$ such that:

\begin{longlist}
\item
\[
P\Biggl(\|\hB'(\hSig{}^{\mathcal{T}}_u)^{-1}\hB-\bB'\Sig_u^{-1}\bB
\|
>Cpm_TK\sqrt{\frac{\log p}{T}}\Biggr)=O\biggl(\frac{1}{p^2}+\frac
{1}{T^2}\biggr);
\]
\item
\[
P\biggl(\|[\hcov(\bff)^{-1}+\hB'(\hSig{}^{\mathcal{T}}_u)^{-1}\hB
]^{-1}\|
>\frac{C}{p}\biggr)=O\biggl(\frac{1}{p^2}+\frac{1}{T^2}\biggr);
\]
\item for $G=[\hcov(\bff)^{-1}+\hB'(\hSig{}^{\mathcal{T}}_u)^{-1}\hB]^{-1}$,
\[
P\bigl(\|\hB G\hB'(\hSig{}^{\mathcal{T}}_u)^{-1}\|>C\bigr)=O
\biggl(\frac
{1}{p^2}+\frac{1}{T^2}\biggr).
\]
\end{longlist}
\end{lem}
\begin{pf}
(i) Let $H=\|\hB'(\hSig{}^{\mathcal{T}}_u)^{-1}\hB-\bB'\Sig
_u^{-1}\bB\|$.
\begin{eqnarray*}
H&\leq&
2\|\bC_T'\Sig_u^{-1}\bB\|+2\bigl\|\bC_T'\bigl((\hSig{}^{\mathcal{T}}_u
)^{-1}-\Sig
_u^{-1}\bigr)\bB\bigr\|\\
&&{}+\bigl\|\bB''\bigl((\hSig{}^{\mathcal{T}}_u)^{-1}-\Sig_u^{-1}\bigr)\bB\bigr\|
+\|\bC _T'\Sig
_u^{-1}\bC_T\|\\
&&{}+\bigl\|\bC_T'\bigl((\hSig{}^{\mathcal{T}}_u)^{-1}-\Sig
_u^{-1}\bigr)\bC_T\bigr\|.
\end{eqnarray*}
The same argument of \citet{FanFanLv08} (equation 14) implies that $\|
{\bB}\|_F=O(\sqrt{p})$. Therefore, by Theorem \ref{t31} and Lemma
\ref
{lb2}(ii), it is straightforward to verify the result.

\mbox{}\hphantom{i}(ii) Since $\|\bD_T\|_F\geq\|\bD_T\|$, according to Lemma \ref
{lb2}(i), there exists $C'>0$ such that with probability ast least
$1-O(T^{-2})$, $\|\bD_T\|<C'K\sqrt{(\log T)/T}$. Thus by Lemma \ref
{la1}, for some $C''>0$,
\begin{eqnarray*}
P\bigl(\|\hcov(\bff_t)^{-1}-\cov(\bff_t)^{-1}\|<C''\|\bD_T\|
\bigr)&\geq&
P\Biggl(\|\bD_T\|<C'K\sqrt{\frac{\log T}{T}}\Biggr)\\
&\geq&1-O(T^{-2}),
\end{eqnarray*}
which implies
%
\begin{equation}\label{eb12}
P\Biggl(\|\hcov(\bff_t)^{-1}-\cov(\bff_t)^{-1}\|<C''C'K\sqrt{\frac
{\log
T}{T}}\Biggr)\geq1-O(T^{-2}).
\end{equation}
Now let $\hA=\hcov(\bff_t)^{-1}+\hB'(\hSig{}^{\mathcal{T}}_u)^{-1}\hB$,
and $\bA=\cov(\bff_t)^{-1}+{\bB}'\Sig_u^{-1}{\bB}$. Then part~(i) and
(\ref{eb12}) imply
%
\begin{eqnarray}\label{eb13}
&&P\Biggl(\|\hA-\bA\|<C''C'K\sqrt{\frac{\log T}{T}}+Cpm_TK\sqrt
{\frac{\log
p}{T}}\Biggr)\nonumber\\[-8pt]\\[-8pt]
&&\qquad\geq1-O\biggl(\frac{1}{p^2}+\frac{1}{T^2}\biggr).\nonumber
\end{eqnarray}
In addition, $m_TK\sqrt{(\log p)/T}=o(1)$. Hence by Lemmas \ref{la1},
\ref{lb4}(ii), for some $C>0$,
\[
P\bigl(\lambda_{\min}(\hA)\geq Cp\bigr)\geq P(\|\hA-\bA\|<Cp)\geq1-O
\biggl(\frac
{1}{p^2}+\frac{1}{T^2}\biggr),
\]
which implies the desired result.

(iii) By the triangular inequality, $\|\hB\|_F\leq\|\bC_T\|
_F+O(\sqrt
{p})$. Hence Lem\-ma~\ref{lb2}(ii) implies, for some $C>0$,
%
\begin{equation}\label{eb14}
P\bigl(\|\hB\|_F\leq C\sqrt{p}\bigr)\geq1-O(T^{-2}+p^{-2}).
\end{equation}
In addition, since $\|\Sig_u^{-1}\|$ is bounded, it then follows from
Theorem \ref{t31} that $\|(\hSig{}^{\mathcal{T}}_u)^{-1}\|$ is bounded
with probability at least $1-O(p^{-2}+T^{-2})$. The result then follows
from the fact that
\[
P(\|G\|>Cp^{-1})=O\biggl(\frac{1}{p^2}+\frac{1}{T^{2}}\biggr),
\]
which is shown in part (ii).
\end{pf}

To complete the proof of Theorem \ref{t32}, part (ii), we follow
similar lines of proof as in \citet{FanFanLv08}. Using the
Sherman--Morrison--Woodbury formula, we have
%
\begin{eqnarray}
&&\|(\hSig{}^{\mathcal{T}})^{-1}-\Sig^{-1}\|\nonumber\\
&&\qquad=\|(\hSig{}^{\mathcal{T}}_u)
^{-1}-\Sig_u^{-1}\|\nonumber\\
&&\qquad\quad{}+\bigl\|\bigl((\hSig{}^{\mathcal{T}}_u)^{-1}-\Sig_u^{-1}\bigr)\hB[\hcov(\bff
)^{-1}+\hB
'(\hSig{}^{\mathcal{T}}_u)^{-1}\hB]^{-1}\hB'(
\hSig{}^{\mathcal{T}}_u)^{-1}\bigr\|
\nonumber\\
&&\qquad\quad{}+\bigl\|\bigl((\hSig{}^{\mathcal{T}}_u)^{-1}-\Sig_u^{-1}\bigr)\hB[\hcov(\bff
)^{-1}+\hB
'(\hSig{}^{\mathcal{T}}_u)^{-1}\hB]^{-1}\hB'\Sig_u^{-1}\bigr\|
\nonumber\\
&&\qquad\quad{}+\|\Sig_u^{-1}(\hB-{\bB})[\hcov(\bff)^{-1}+\hB'(
\hSig{}^{\mathcal{T}}_u)^{-1}\hB]^{-1}\hB'\Sig_u^{-1}\|
\\
&&\qquad\quad{}+\|\Sig_u^{-1}(\hB-{\bB})[\hcov(\bff)^{-1}+\hB'(
\hSig{}^{\mathcal{T}}_u)^{-1}\hB]^{-1}{\bB}'\Sig_u^{-1}\|
\nonumber\\
&&\qquad\quad{}+\bigl\|\Sig_u^{-1}{\bB}\bigl([\hcov(\bff)^{-1}+\hB'(\hSig{}^{\mathcal{T}}_u
)^{-1}\hB]^{-1}\nonumber\\
&&\qquad\quad\hspace*{51.6pt}{}-[{\cov}(\bff)^{-1}+{\bB}'\Sig_u^{-1}{\bB
}]^{-1}\bigr){\bB
}'\Sig_u^{-1}\bigr\|
\nonumber\\
&&\qquad=L_1+L_2+L_3+L_4+L_5+L_6.\nonumber
\end{eqnarray}

The bound of $L_1$ is given in Theorem \ref{t31}.

For $G=[\hcov(\bff)^{-1}+\hB'(\hSig{}^{\mathcal{T}}_u)^{-1}\hB
]^{-1}$, then
%
\begin{equation}\label{eb7}
L_2\leq\|(\hSig{}^{\mathcal{T}}_u)^{-1}-\Sig_u^{-1}\|\cdot\|\hB G\hB
'(\hSig{}^{\mathcal{T}}_u)^{-1}\|.
\end{equation}
It follows from Theorem \ref{t31} and Lemma \ref{lb5}(iii) that
\[
P\Biggl(L_2\leq Cm_TK\sqrt{\frac{\log p}{T}}\Biggr)\geq1-O
\biggl(\frac
{1}{p^2}+\frac{1}{T^{2}}\biggr).
\]

The same bound can be achieved in a same way for $L_3$. For $L_4$, we have
\[
L_4\leq\|\Sig_u^{-1}\|^2\cdot\|\hB-\bB\|\cdot\|\hB\|\cdot\|G\|.
\]
It follows from Lemmas \ref{lb2}, \ref{lb5}(ii), and inequality (\ref
{eb14}) that
\[
P\Biggl(L_4\leq C\sqrt{\frac{K\log p}{T}}\Biggr)\geq1-O\biggl(\frac
{1}{p^2}+\frac{1}{T^2}\biggr).
\]
The same bound also applies to $L_5$. Finally,
\[
L_6\leq\|\bB\|^2\|\Sig_u^{-1}\|^2\|\hA^{-1}-\bA^{-1}\|\leq\|\bB\|
^2\|
\Sig_u^{-1}\|^2\|\hA-\bA\|\cdot\|\hA^{-1}\|\cdot\|\bA^{-1}\|,
\]
where both $\hA$ and $\bA$ are defined after inequality (\ref{eb12}).
By Lemma \ref{lb4}(ii), $\|\bA^{-1}\|=O(p^{-1})$. Lemma \ref{lb5}(ii)
implies $
P(\|\hA^{-1}\|>Cp^{-1})=O(p^{-2}+T^{-2})$. Combining with (\ref{eb13}),
we obtain
\[
P\Biggl(L_6\leq Cm_TK\sqrt{\frac{\log p}{T}}\Biggr)\geq1-O
\biggl(\frac
{1}{p^2}+\frac{1}{T^{2}}\biggr).
\]
The proof is completed by combining $L_1\sim L_6$.

\section{\texorpdfstring{Proofs for Section \lowercase{\protect\ref{sec4}}}{Proofs for Section 4}}\label{appC}

The proof is similar to that of Lemma \ref{l31}. Thus we sketch it
very briefly. The OLS is given by
\[
\widehat{\mathbf b}_i=(\bX_i'\bX_i)^{-1}\bX_i'\by_i,\qquad i\leq p.
\]
The same arguments in the proof of Lemma \ref{lb1} can yield, for large
enough $C>0$,
\[
P\Biggl(\max_{i\leq p} \|\widehat{\mathbf b}_i-\mathbf{b}_i\|
>C\sqrt{\frac{K\log p}{T}}
\Biggr)=O\biggl(\frac{1}{p^2}+\frac{1}{T^2}\biggr),
\]
which then implies the rate of
\[
\max_{i\leq p}\frac{1}{T}\sum_{t=1}^T(u_{it}-\hat{u}_{it})^2\leq
\max
_{i\leq p}\|\widehat{\mathbf b}_i-\mathbf{b}_i\|^2\frac
{1}{T}\sum_{t=1}^T\|\bff_{it}\|^2.
\]
The result then follows from a straightforward application of Theorem
\ref{t21}.
\end{appendix}



\printaddresses

\end{document}